\documentclass[preprints,article,accept,moreauthors,pdftex]{Definitions/mdpi} 

\firstpage{1} 
\pubvolume{1}
\issuenum{1}
\articlenumber{0}
\pubyear{2021}
\copyrightyear{2020}

\pdfoutput=1

\usepackage{subcaption}

\Title{A Novel Manufacturing Process for Glass THGEMs and First Characterisation in an Optical Gaseous Argon TPC}

\Author{Adam Lowe, Krishanu Majumdar, Konstantinos Mavrokoridis *, Barney Philippou *, Adam Roberts * and Christos Touramanis}

\AuthorNames{Barney Philippou, Konstantinos Mavrokoridis and Adam Roberts}

\address{University of Liverpool, Department of Physics, Oliver Lodge Bld, Oxford Street, Liverpool, L69 7ZE, UK}

\corres{Correspondence: k.mavrokoridis@liverpool.ac.uk (K.Mv.); sgbphili@liverpool.ac.uk (B.P); aroberts@hep.ph.liv.ac.uk (A.R.) }

\abstract{This paper details a novel, patent pending, abrasive machining manufacturing process for the formation of sub-millimetre holes in THGEMs, with the intended application in gaseous and dual-phase TPCs. Abrasive machining favours a non-ductile substrate such as glasses or ceramics. This innovative manufacturing process allows for unprecedented versatility in THGEM substrates, electrodes, and hole geometry and pattern. Consequently, THGEMs produced via abrasive machining can be tailored for specific properties, for example: high stiffness, low total thickness variation, radiopurity, moisture absorption/outgassing and/or carbonisation resistance. This paper specifically focuses on three glass substrate THGEMs (G-THGEMs) made from Schott Borofloat 33 and Fused Silica. Circular and hexagonal hole shapes are also investigated. The G-THGEM electrodes are made from Indium Tin Oxide (ITO), with a resistivity of 150~$\Omega$/Sq. All G-THGEMs were characterised in an optical (EMCCD) readout GArTPC, and compared to a traditionally manufactured FR4 THGEM, with their charging and secondary scintillation (S2) light production behaviour analysed. }


\begin{document}
\setcounter{secnumdepth}{2}

\begin{center}
\includegraphics[width=0.35\textwidth]{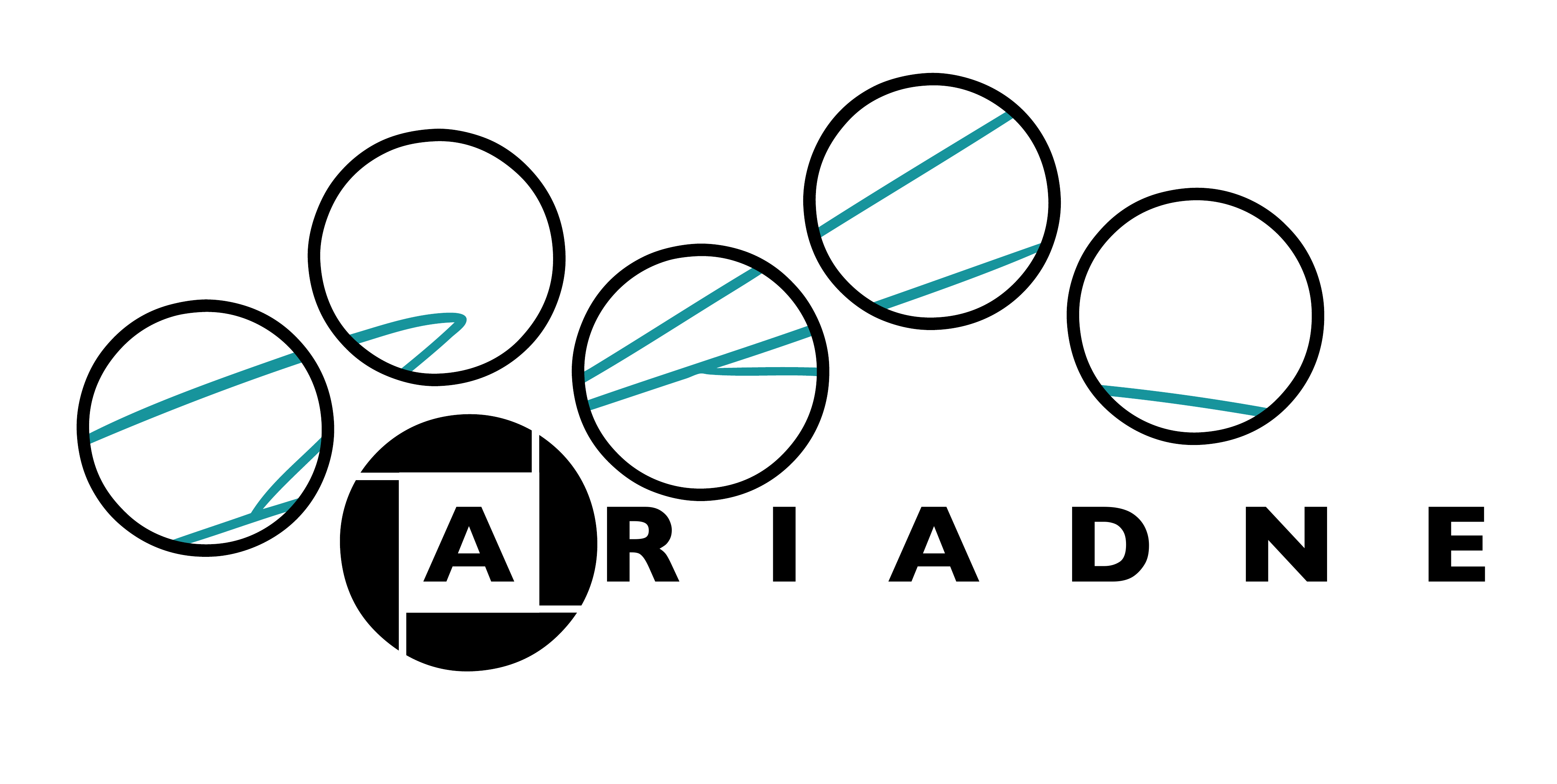}
\end{center}


\section{Introduction} \label{Introduction}
Recent years have seen significant advancements in Micro-Pattern Gaseous Detectors (MPGDs). Gaseous Electron Multipliers (GEMs) \cite{GEM} and THick Gaseous Electron Multipliers (THGEMs) (also know as Large Electron Multipliers (LEMs)) \cite{THGEM} are some of the most recent developments within the field of MPGDs, and have proved important for their simplicity and effectiveness. GEMs and THGEMs both have similar basic arrangements: an insulating substrate, sandwiched between two electrically conductive electrodes, with holes perforating through all three layers. Compared to GEMs, THGEMs typically have an order of magnitude increased thickness. Traditionally, THGEMs are manufactured from an epoxy laminate/FR4 substrate and have copper plated electrodes. Sub-millimetre through-holes are mechanically drilled through the device to form field amplification regions. A dielectric rim may be etched around each hole. It has been reported that dielectric rims reduce the probability of discharges, ultimately resulting in ten-fold higher gain compared to designs without rims \cite{THGEMDetectors}. 

The robustness and simplicity of THGEMs, both in terms of their use and manufacturing, has resulted in their adoption in a wide variety of applications. This paper will concentrate on their use in dual-phase and gaseous Time Project Chambers (TPCs). Typically, TPCs can either be filled completely with a scintillating gas or, in the dual-phase case, filled with both the liquid and gaseous state of a single scintillator (typically argon or xenon). When an ionising particle passes through the TPC volume, both scintillation light and free electrons are produced. By means of an applied electric field, the free electrons are drifted to the surface of the TPC, which is typically covered by devices capable of detecting these electrons. By detecting both the scintillation light and the ionised electron signal, event reconstruction within the TPC is possible. In the case of gaseous TPCs, both initial scintillation/ionisation and detection are performed in the gas. In dual-phase TPCs, initial scintillation/ionisation occurs in the liquid, affording the benefit of increased target density. The ionised signal is drifted and extracted into the gaseous phase for detection. The benefit of extraction into a gaseous phase is retaining the ability to amplify the ionised signal using MPGDs, which typically offer favourable performance when operating in gas compared to liquid. The effectiveness of THGEMs has resulted in widespread use within the contemporary TPC field, and much development has been made including the production of Liquid Hole Multipliers and FAT GEMs \cite{Buzulutskov2020_ReviewChargeAmplification}. TPCs are now integral within the Neutrino and Dark Matter sectors, with the proposed kiloton scale DUNE modules \cite{DuneCDRVol1, DuneCDRVol2, DuneCDRVol3, DuneCDRVol4, DuneIDRVol1, DuneIDRVol2, DuneIDRVol3}, CYGNO \cite{CYGNO}, DarkSide-20k \cite{Darkside}, ArDM \cite{ArDM} and LZ \cite{LZ}. 

Despite the many attractive features of THGEMs, there are limitations in terms of their design and manufacture. Conventionally, THGEMs may be manufactured using standard printed circuit board (PCB) techniques. This typically results in THGEMs produced from epoxy laminate/FR4 substrates. Epoxy laminate/FR4 typically contains radioactive contaminants, precluding use within the dark matter community. Additionally, epoxy laminate/FR4 is a porous material, hence contamination, moisture absorption and outgassing issues may occur. Large area THGEMs/LEMs typically require mechanical support within the active area to limit deformations caused by sagging under gravity. Variations in epoxy laminate/FR4 thickness across the substrate may also result in variations in field (and therefore gain) across the THGEM. The use of substrates with higher stiffness and stricter thickness tolerances may alleviate both of these issues. Repeated sparking and discharging of THGEMs during operation may eventually lead to carbonisation of the epoxy laminate/FR4 substrates. Carbonisation forms a conductive pathway between electrodes, leading to degraded THGEM performance over time, potentially resulting in device failure. Since THGEM holes are traditionally formed via mechanical drilling, they maybe subject to variations in hole size across the THGEM due to drill bit wear. This may lead to further gain non-uniformities. 

This paper details the manufacture and characterisation of Glass THGEMs (G-THGEMs), created using a novel masked abrasive machining process (patent pending \cite{Mavrokoridis2020_G-THGEM}), within the ARIADNE R\&D framework \cite{ARIADNE_TDR}. Performance of the G-THGEMs is compared to a conventionally manufactured THGEM. The conventional THGEM discussed in this paper has a 1~mm thick FR4 substrate and copper-coated electrodes, with 500~$\mu$m diameter holes formed via mechanical drilling on an 800~$\mu$m pitch hexagonal array, with 50~$\mu$m etched rims. 

Glass GEMs (G-GEMs), produced from photosensitive glass, have already been fabricated (and tested), via a photolithography technique \cite{Takahashi_2013, Fujiwara_2019}. The new manufacturing technique presented in this paper is not limited to photosensitive glass and is suitable for a wide range of substrates. Both substrate and electrode materials can be tailored depending on specific application requirements - for example, high stiffness, low outgassing and/or high radiopurity. Abrasive machining allows for unprecedented versatility in THGEM patterning in terms of hole shape and layout. In addition, the abrasive machining process is no longer subject to drill bit wear, potentially improving hole diameter consistency across the THGEM. The abrasive machining manufacturing process could have far reaching consequences within the Neutrino and Dark Matter sectors \cite{CYGNO, Darkside, ArDM, LZ, Gai_2007_DarkMatterTHGEMs}, as well as in Medical Imaging \cite{Tsyganov_2008_Medical}.
 
\section{Novel G-THGEMs Manufacturing Process}
\label{ManufacturingProcess}

\begin{figure}[h!]
    \centering
    \includegraphics[width=0.7\textwidth]{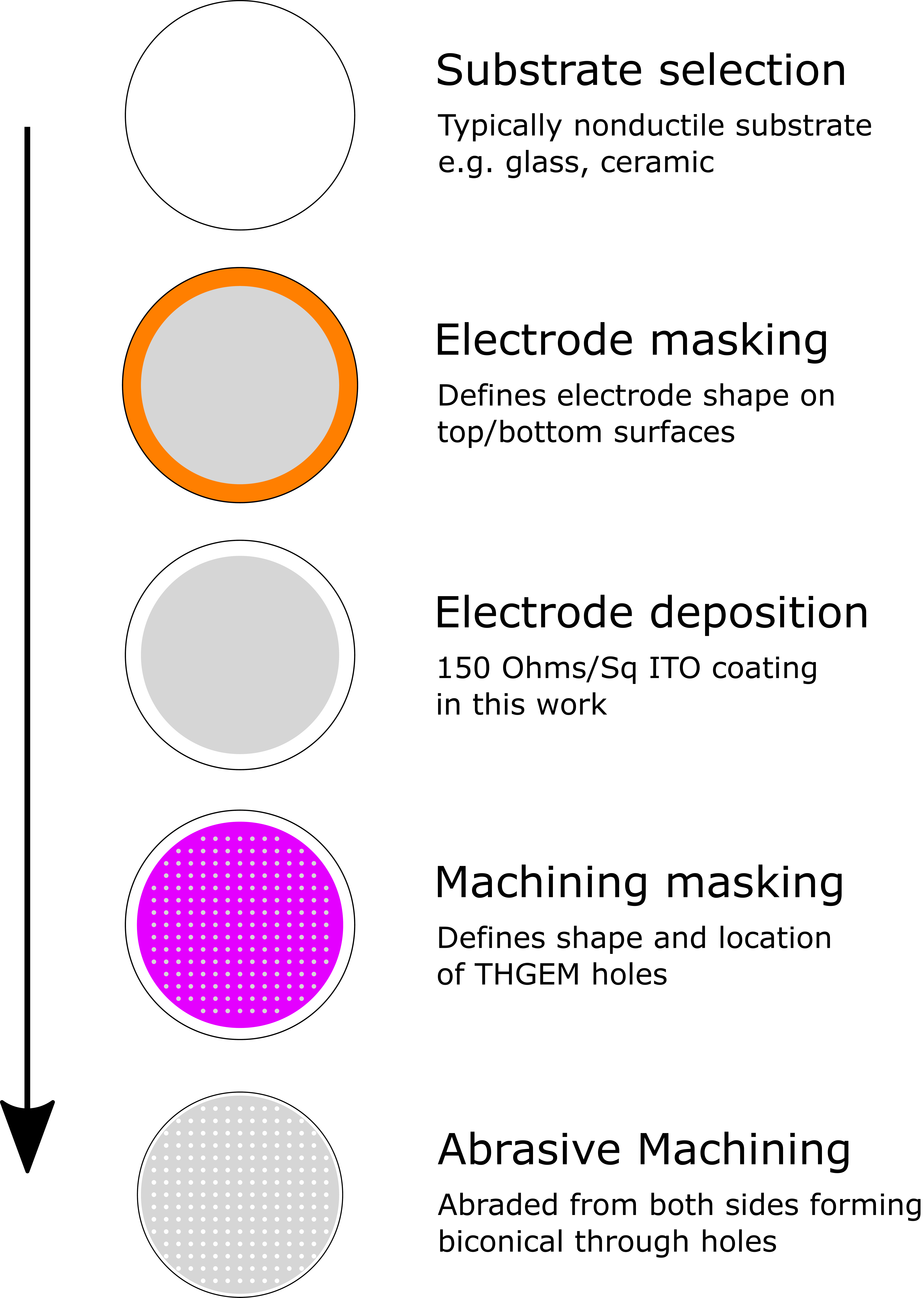}
    \caption{A visual representation of the novel, patent pending, abrasive machining technique for the production of G-THGEMs.}
    \label{fig:GlassTHGEMProduction}
\end{figure}

The manufacturing process for G-THGEMs is composed of several distinct steps. A visual representation is shown in Figure~\ref{fig:GlassTHGEMProduction}. First, a substrate is selected. The substrate must generally be nonductile, favoring materials such as glass or ceramics. In this work, Fused Silica and Schott Borofloat 33 substrates were tested. The exact substrate can be tailored according the requirements of the application. For example, synthetic Fused Silica can be made especially radiopure \cite{Abgrall2016} and would therefore be ideally suited for experiments requiring low backgrounds. Schott Borofloat 33 is relatively lower cost and may be well suited for applications requiring high resistance to thermal shock. For this investigation, all substrates were selected to be 1~mm thick, identical to that of the traditional FR4 THGEM.

\begin{figure} [h!]
    \centering
    \includegraphics[width=0.8\textwidth]{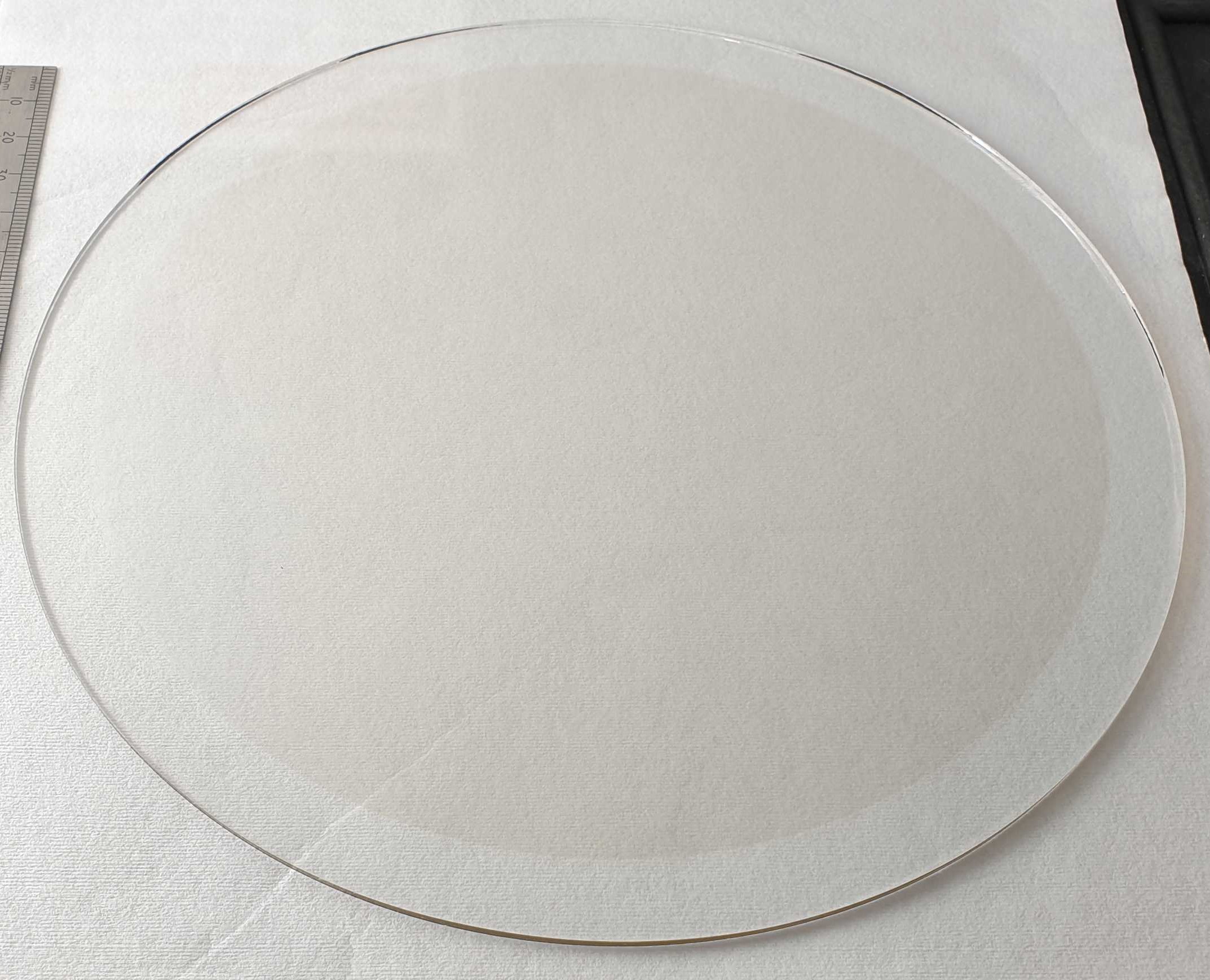}
    \caption{A 1~mm thick, 200~mm diameter glass wafer, coated with a 163~mm diameter Indium Tin Oxide (ITO) electrode on both the top and bottom faces.}
    \label{fig:ITOGlass}
\end{figure}

The next step in the manufacturing process is the formation of electrodes. In this work, a vacuum deposited 150~$\Omega$/Sq Indium Tin Oxide (ITO) coating is applied on both sides of the substrate. A simple mask, as shown in orange on Figure~\ref{fig:GlassTHGEMProduction}, is used during the coating process to define the shape of the electrode - in this case, the electrode area is a 163~mm diameter circle on the 200~mm diameter substrates. A photograph of the substrate with ITO coating is shown in Figure~\ref{fig:ITOGlass}. ITO was selected predominantly for its relatively low cost and good availability, although many different electrode materials could be applied. The controllable sheet resistance of the ITO coating may provide additional benefits when compared to highly conductive coatings - this is discussed further in Section~\ref{GEM Types}. More intricate electrode geometries may also be possible via laser etching of the electrode coating \cite{LaserEtching}.  

\begin{figure} [h!]
\centering
\includegraphics[width=0.9\textwidth]{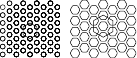}
\caption{The hole topographies for circular and hexagonal masks, with the repeat unit overlayed. These masks dictate the hole structure of the G-THGEMs, formed via abrasive machining. The hole sizes and pattern dimensions are given in Table~\ref{tab:THGEMTypes}.} 
\label{fig:BlastingMask}
\end{figure}

The substrate, now with electrodes on both sides, is ready for the through hole machining process. The position and shape of the through holes is defined using a 2D CAD drawing, such as Drawing Exchange Format (DXF) file. For best results during the machining process, the diameter of the holes should typically be at least half of the substrate thickness. For example, for a 1~mm thick substrate, hole diameters $\geq 0.5$~mm are preferred. Figure~\ref{fig:BlastingMask} shows a schematic of two hole layouts which were tested in this work. One design uses the conventional THGEM dimensions: 0.5~mm diameter holes on a hexagonal array with a hole-to-hole pitch of 0.8~mm. The second design resembles a honeycomb structure, with hexagonal holes on a hexagonal array. This design highlights new capabilities provided by the manufacturing process - no longer being limited to geometries which can be produced by drilling.

\begin{figure} [h!]
    \centering
    \includegraphics[width=0.9\textwidth]{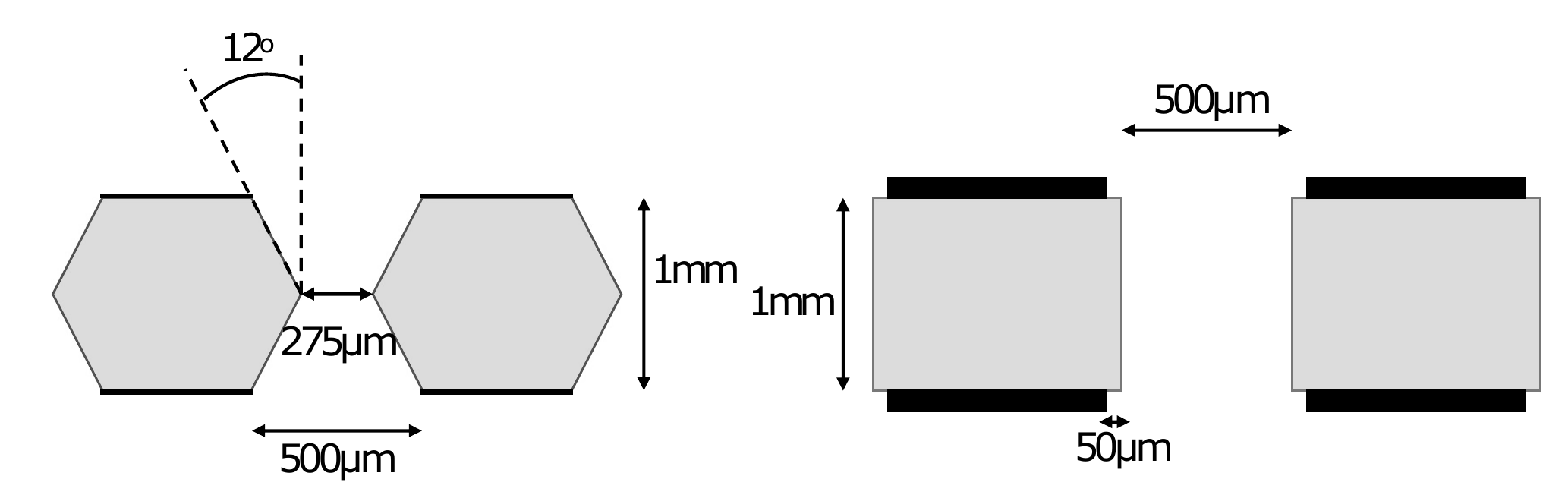}
    \caption{A cutaway schematic of the hole geometry for the G-THGEM (left), formed by abrasive machining, and the mechanically drilled holes of a conventional THGEM (right). The G-THGEM/THGEM substrate are shown in grey, with electrodes represented in black. The G-THGEMs are rimless, with a bi-conical shaped hole, while the THGEM has a 50~$\mu$m rim and straight walled holes.}
    \label{fig:HoleShapesDiagram}
\end{figure}

\begin{figure} [h!]
\centering
\includegraphics[width=0.9\textwidth]{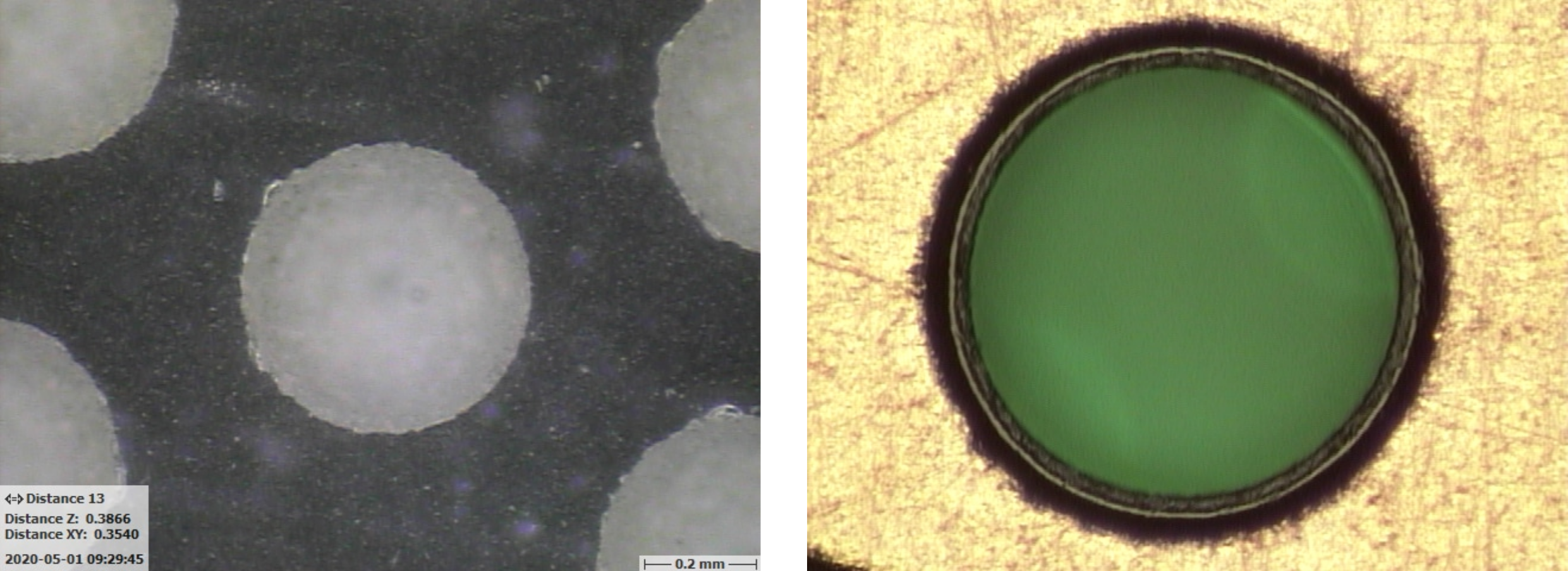}
\caption{A comparison between the abrasively machined holes in a Borofloat 33 G-THGEM (left) and the drilled holes in the FR4 THGEM (right). Both G-THGEM/THGEM holes are 500~$\mu$m in diameter.} 
\label{fig:THGEMHoleComparison}
\end{figure}

The resulting mask, shown in purple on Figure~\ref{fig:GlassTHGEMProduction}, which defines location and shape of the through holes, is applied to both sides of the substrate. This mask is patterned via a photolithography technique, selectively producing areas which resist the abrasive machining process. Then, a series of abrasive delivery nozzles traverse the substrate, selectively abrading the substrate in the regions which are not masked. Typical abrasive materials includes aluminium oxide or silicon carbide, with a particle size of around 20-25~$\mu$m. When performed from one side, the abrasion process produces holes which taper inwards with increasing depth into the substrate. The taper angle (controllable between 6 - 35~degrees) was measured to be approximately 12~degrees in this work. Once the process is completed on one side of the substrate, the substrate is flipped and the process is repeated on the other side. By abrading the substrate from both sides, bi-conical through holes are produced. The outer perimeter of the substrate can also be shaped by this process. In this work, the diameter was reduced to 179~mm. A schematic comparing holes of the G-THGEM to those of a traditional FR4 THGEMs is shown in Figure~\ref{fig:HoleShapesDiagram}. A photograph comparing the G-THGEM and THGEM holes can be seen in Figure~\ref{fig:THGEMHoleComparison}. Small edge chipping (less than 10~$\mu m$) occasionally occurs around the G-THGEM holes from the abrasive machining process. No discernible affects on the behaviour of the G-THGEMs were identified, including discharge behaviour around these holes. Optimisation of the production technique is ongoing including varying the abrasive media, particulate size and the delivery pressure.  The bi-conical shape of G-THGEM holes produces distinctive dielectric charging behaviour during operation - this is discussed further in Section~\ref{characterisation}. 

\begin{figure} [h!]
    \centering
    \includegraphics[width=0.8\textwidth]{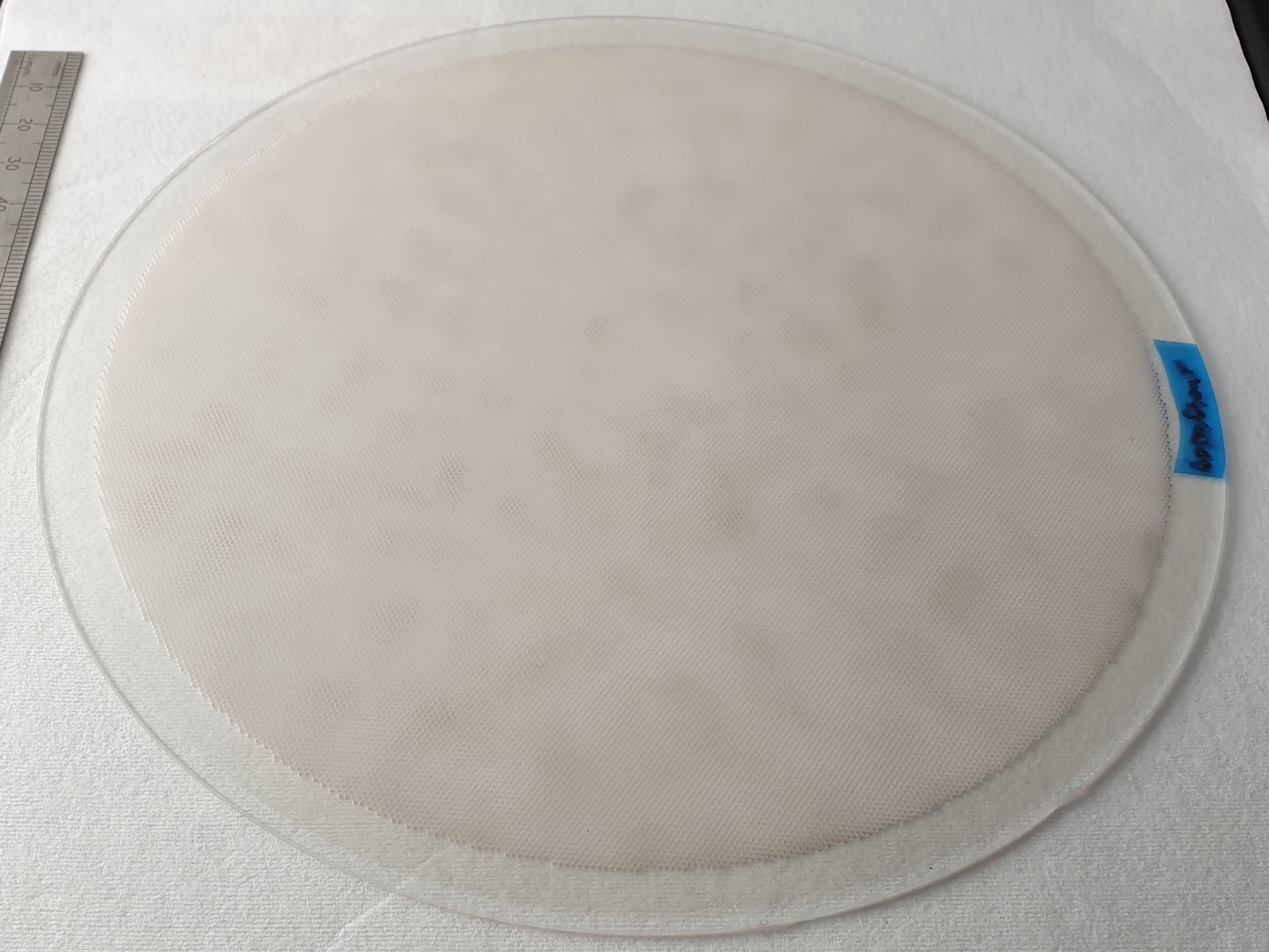}
    \caption{A Borofloat 33 substrate G-THGEM, coated on the top and bottom face with an ITO resistive coating, with micro-patterned hole formation created via a masked abrasive machining process.}
    \label{fig:GlassTHGEM}
\end{figure}

Following the abrasive blasting process, the mask is removed and the final G-THGEM is realised. A photograph of a complete G-THGEM is shown in Figure~\ref{fig:GlassTHGEM}. For these G-THGEM prototypes no special cleaning processes were afforded before use. However, multiple types of cleaning could be permitted, for example, the use of an ultrasonic bath. In this work, electrical connection is made to the top and bottom ITO electrodes using spring loaded contacts connected to the the outer perimeter of the electrode.

\section{Experimental Setup} \label{ExperimentalSetup}

\subsection{G-THGEMs}
\label{GEM Types}
 Table~\ref{tab:THGEMTypes} summarises the four THGEMs, with their respective combinations of substrate material, electrode coating and hole shape and size. Both the G-THGEMs and FR4 THGEM have a 179~mm diameter, with a 163~mm diameter active area. 

Two substrate materials were tested for the G-THGEMs: SCHOTT Borofloat 33 \cite{SCHOTT_BF, borofloatData, uqgoptics, SCHOTT_TechnicalGlasses} and Fused silica \cite{SCHOTT_FS, fusedSilicaData, SCHOTT_TechnicalGlasses}. Borofloat 33 is relatively low cost, enabling potential large scale production of relatively inexpensive G-THGEMs, whereas Fused Silica is radiopure, potentially allowing use in experiments which are very sensitive to background radiation (for example, dark matter TPCs).

The abrasive machining process allows for more complex THGEM hole topography than conventional manufacturing methods. To highlight this capability, a G-THGEM with hexagonal holes was also tested. Hexagonal holes can be packed more densely (shown in Figure~\ref{fig:BlastingMask}), offering higher open areas and possibly increased electron transparency. The open area of the hexagonal G-THGEM is 62~$\%$, compared to 35~$\%$ of the circular holes.

\begin{table}[h!] 
\centering
    \small
        \setlength\tabcolsep{2pt}
        \centerline
        
    \begin{tabular}{|c|c|c|c|c|c|c|}
        \cline{1-7}
        \textbf{THGEM Type} & \textbf{Substrate} & \textbf{Electrode} & \textbf{Hole Shape} & \textbf{Hole Size ($\mu$m)} & \textbf{Hole Pitch ($\mu$m)} & \textbf{Rim Size ($\mu$m)} \\
    \hline
        FR4 circ & FR4 & Cu & Circle & 500 & 800 & 50\\
        FS circ & FS & ITO & Circle & 500 & 800 & None \\
        BF circ & BF & ITO & Circle & 500 & 800 & None \\
        BF hex & BF & ITO & Hexagonal & 870 & 1100 & None \\
        \cline{1-7}
    \end{tabular}
    
    \caption{A summary of the G-THGEM/THGEM types investigated in this paper, including substrates, electrodes, hole pattern, size, shape and rims. The short-hand naming convention represents the G-THGEM/THGEM's substrate and hole shape, where "FS" and "BF" refers to Fused Silica and Schott Borofloat 33, respectively.}
    \label{tab:THGEMTypes}
\end{table}

There are several mechanical advantages afforded by G-THGEMs which makes them suitable for large-scale LArTPC experiments. Firstly, both Borofloat 33 and Fused Silica are stiffer than FR4 and can be produced in large flat surfaces, with thickness variation tolerances typically superior than what is possible for FR4. Large-scale experiments, for example DUNE, utilise PCBs substrate LEMs with reported thickness variations approaching 5~$\%$, with a rejection rate of less than 1 LEM per Charge Readout Plane (CRP) (18 LEMs) \cite{DuneIDRVol3}. Borofloat 33 sheets are available with thickness variations of $\leq15~\mu m$ for 1~mm thick substrates \cite{SCHOTT_GlassWafers}, resulting in a thickness variation of less than $1.5\%$. This is an important consideration in achieving high THGEM gain uniformity. Differential surface flatness and THGEM bowing both contribute to THGEM field distortions. SCHOTT Borofloat 33 and Fused Silica are reported with flexural moduli of 64~GPa \cite{SCHOTT_BF, borofloatData} and 72~GPa \cite{SCHOTT_FS, fusedSilicaData} respectively. The flexural strength of epoxy laminate/FR4 used to produce THGEMs does not typically exceed 23~GPa (crosswise) \cite{Isola_DE104, Eltos_Materials}.

Glass is less porous than FR4, making it less susceptible to contamination. This presents obvious advantages to detectors with high purity requirements. A common type of G10/FR4 is Textolite G 10 FR4, manufactured by General Electric, and this has been reported to have a Total Mass Loss (TML) and Collected Volatile Condensable Material (CVCM) outgassing rate of $0.44\%$ and $0.01\%$, respectively \cite{ESA}. Silica based glasses/ceramics such as Borosilicate (Borofloat), Fused Silica or Aluminium Silicate typically have relatively lower outgassing rates \cite{ESA}.

\subsection{The ARIADNE prototype TPC}
\label{ARIADNE}
\begin{figure} 
    \centering
    \includegraphics[width=1.0\textwidth]{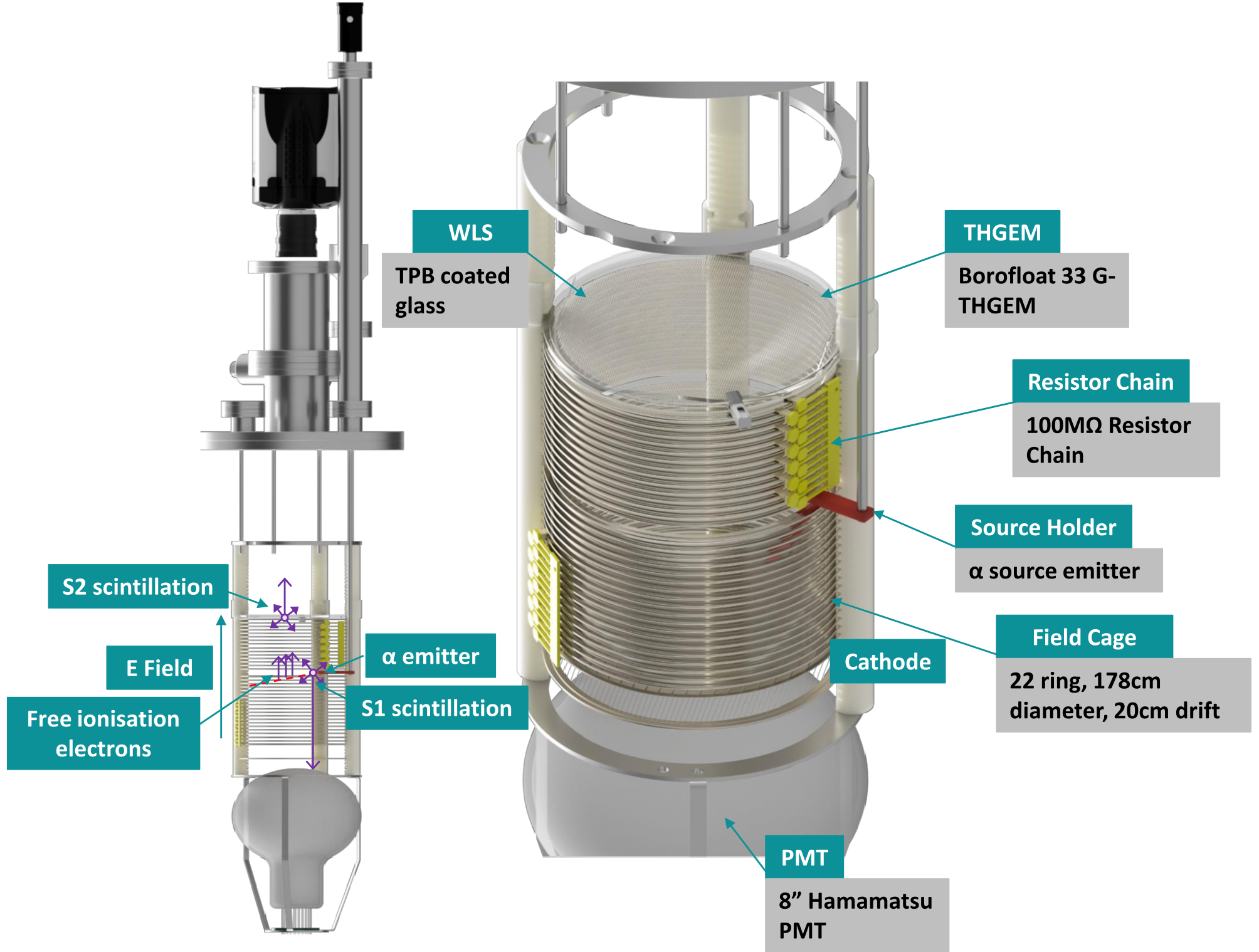}
    \caption{The ARIADNE Prototype (right) and operating principle (left). As a single phase, optical readout GArTPC, a wavelength shifter (TPB) is used to shift the 128~nm VUV S2 scintillation light to 420~nm visible light. This is then imaged with an EMCCD.}
    \label{fig:LARTPC_drift_rotated_label}
\end{figure}

Each of the four G-THGEMs/THGEM were characterised using the ARIADNE prototype detector. The ARIADNE prototype \cite{ARIADNE_Prototype}, shown in Figure~\ref{fig:LARTPC_drift_rotated_label}, is an optical readout, 40~L cylindrical TPC, with both dual-phase cryogenic and gas capabilities. Characterisation of all THGEMs was undertaken in GAr conditions.

The TPC field cage, comprised of  22 stainless steel rings, has a 178mm diameter with a 20~cm drift length. The field cage is bounded by a cathode grid at the bottom of the TPC and capped at the top with the bottom electrode of the THGEM. The drift field is established between the cathode and bottom THGEM electrode. Outside the TPC active region, mounted on a rotatory feedthrough, is an Americium-241 alpha source, with an initial 30~kBq rate. This is collimated to around 500~Bq. The rotary feedthrough allows the source to be rotated in/out of the TPC volume during operation.

A sheet of Tetraphenyl Butadiene (TPB) coated wavelength shifting (WLS) acrylic is mounted directly above the THGEM. This WLS plate shifts the VUV S2 scintillation light from 128~nm \cite{ArgonScintillation} to 420~nm \cite{TPBSpectrum, Benson2017_TPBSpectrum}. This wavelength is in the high quantum efficiency range of the externally mounted Andor iXon Ultra 888 EMCCD camera \cite{AndorDataSheet} (further discussed in \cite{ARIADNE_TDR}). The Andor iXon Ultra 888 EMCCD camera is positioned externally to the detector, at a distance of approximately 600~mm from the THGEM. The EMCCD is combined with a Spacecom VF50095M lens, with a speed of f/0.95, and a focal length of 50~mm, giving a field of view of approximately 160x160~mm$^2$. The S2 scintillation light is imaged through a 90~mm diameter Borosilicate glass viewport. The transparency of the viewport is $> 90~\%$ \cite{ViewportDataSheet} at 420~nm (the peak of the TPB emission spectrum). The camera is mounted on an adjustable tripod frame which sits on the optical viewport. It was important to maintain consistent field of view and focusing between studies. This was achieved through reinforcement of the legs with locking nuts and retaining EMCCD orientation during reassembly.

An 8-inch, TPB coated, Hamamatsu R5912-20-MOD  Photomultipler Tube (PMT) is mounted below the TPC. The spectral response range of the PMT lies between 300 and 650~nm, with a quantum efficiency of approximately $30~\%$ at 420~nm \cite{PMTDataSheet}. 
 
\begin{figure}
\centering
\begin{subfigure}{.4\textwidth}
  \centering
   \includegraphics[width=0.9\textwidth]{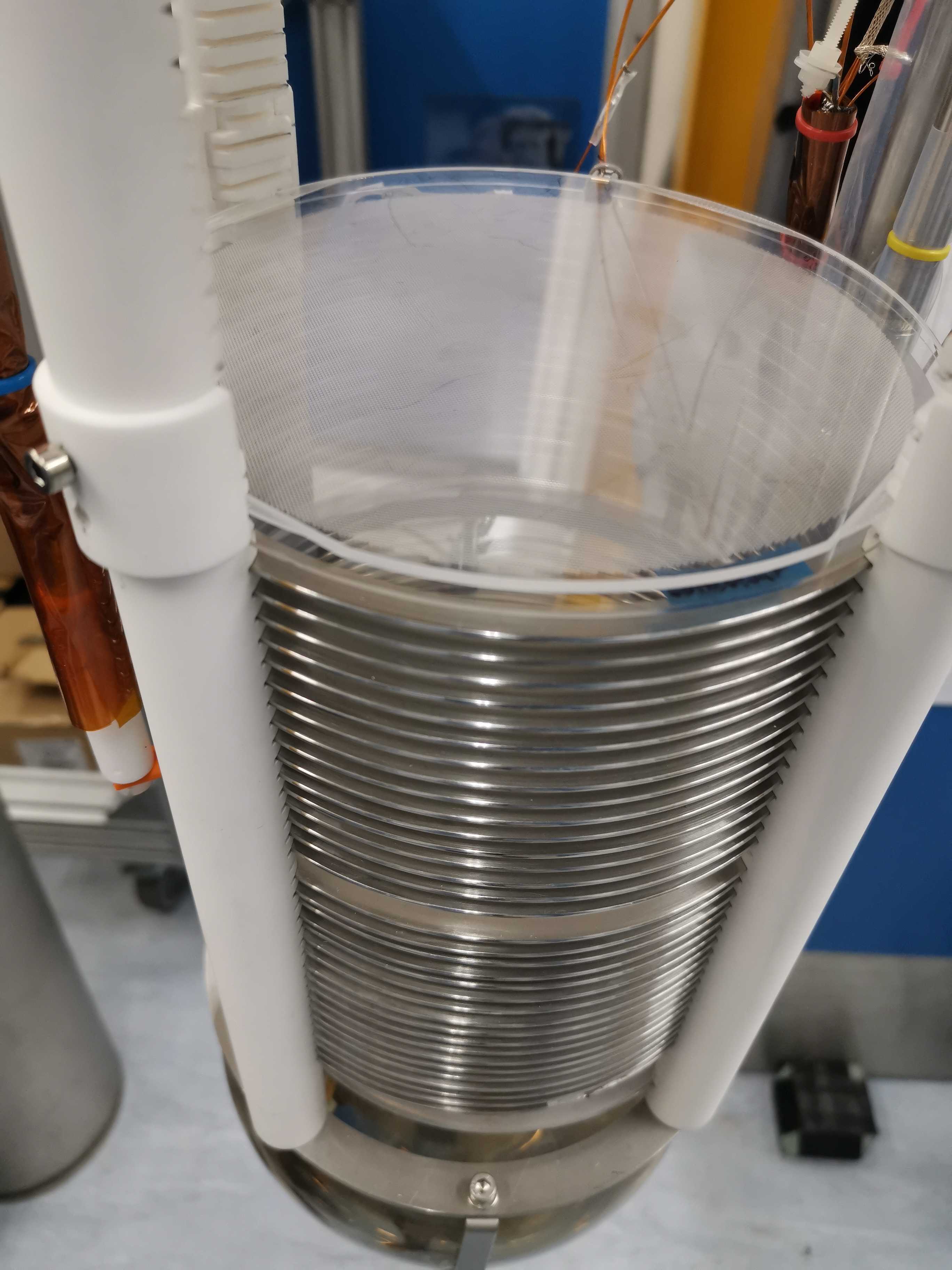}
  \caption{Borofloat G-THGEM.}
  \label{fig:PrototypeBF}
\end{subfigure} %
\begin{subfigure}{.4\textwidth}
  \centering
   \includegraphics[width=0.9\textwidth]{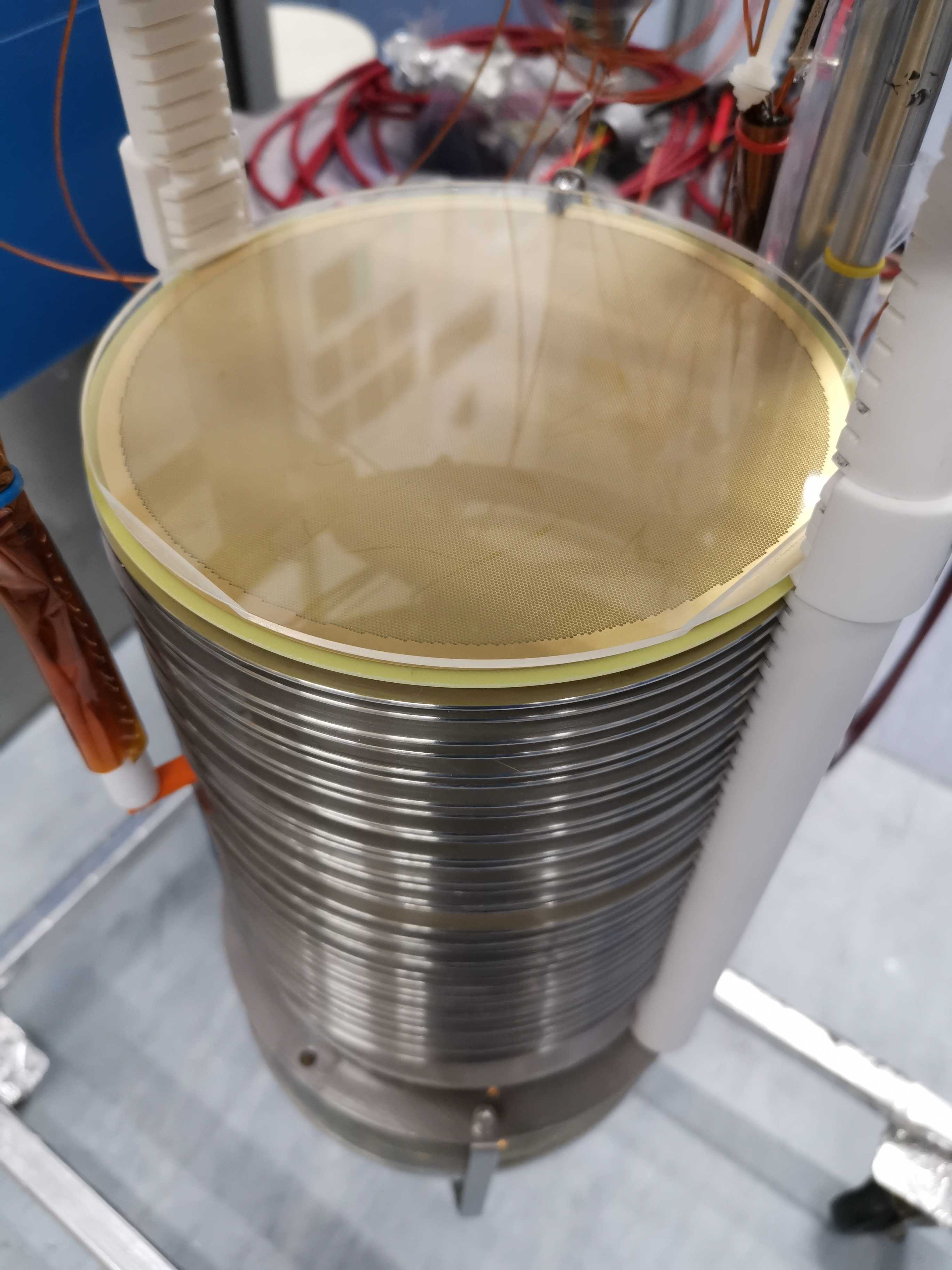}
  \caption{FR4 THGEM.}
    \label{fig:PrototypeGold}
\end{subfigure}
\caption{A Borofloat 33 G-THGEM and conventional copper-coated FR4 PCB THGEM mounted on the ARIADNE prototype TPC.}
 \label{fig:PrototypeTHGEMs}
\end{figure}

\subsection{TPC Operation Principle}
The operating principle of the optical readout and ARIADNE prototype TPC in GAr conditions is shown on the left of Figure~\ref{fig:LARTPC_drift_rotated_label}. All G-THGEM/THGEM behaviours were characterised by monitoring the intensity of the S2 light produced in the THGEM holes. The primary scintillation/ionisation was induced by alpha particle/GAr interactions. The emitted S2 light was imaged with the EMCCD.

An emitted alpha particle enters the TPC active region and interacts with GAr, resulting in emission of prompt scintillation light (S1), and the creation of free electrons by ionisation. The S1 light is detected by the PMT at the bottom of the detector and is used for $\tau_2$ purity monitoring, which is further discussed in Section~\ref{Purity}. 

Free ionisation electrons drift upwards in response to a uniform electric field applied between the cathode and the grounded G-THGEM/THGEM bottom electrode. When electrons reach the top of the field cage, they enter the G-THGEM/THGEM holes. Each electron is then accelerated by the high electric field applied across the THGEM, between bottom and top electrodes. At sufficiently high fields, this acceleration causes the electron to further ionise the GAr resulting in a Townsend discharge. This exponentially increases the number of electrons. In addition, this process also produces secondary scintillation light - the S2 light signal \cite{THGEMReadout, THGEMLightYield}, through excitation of GAr atoms. Depending on applied field, on the order of 100s of photons are produced per accelerated electron \cite{THGEMReview_Buzulu}. The EMCCD camera images this S2 light, which has been wavelength shifted using TPB to the high quantum efficiency range of the camera. 

\subsection{Experimental Procedure} \label{Experiment Procedure}
The TPC chamber was evacuated over 24~hours, down to rough vacuum of $10^{-3}$~mbar.  The chamber was then filled to $1100$~mbar with N6.0 GAr ($99.9999~\%$ pure). GAr was continuously flushed through the detector volume, at flow rate of approximately 5~L/min for the duration of the THGEM studies, mitigating any outgassing effects that would otherwise degrade purity. Purity, measured using the slow scintillation component - $\tau_2$ lifetime, was monitored throughout, as described in Section~\ref{Purity}. 

EMCCD camera settings were kept consistent throughout the duration of the experiment - operating with 4x4 binning, in full frame mode, with an exposure time of 50~ms. The EMCCD gain value was set its maximum value of 1000. In this configuration, the EMCCDs had a readout rate of 15~Hz. This rate resulted in approximately 30 alpha GAr interactions recorded in a single image. The cameras were air cooled to less than -60~Celsius before recording to minimise sensor noise.

The TPC drift field was established between the cathode, operated at a bias of -2.5~kV, and the G-THGEM/THGEM bottom electrode, which was grounded. All field shaping rings were connected with a 100~M$\Omega$ resistor chain.  The THGEMs were left unbiased and grounded for a 24~hour period (in order to be fully discharged) before measurements were taken. Characterisation began with a study into the dielectric charging behaviour of the G-THGEMs/THGEM. During measurements, biases were applied to the detector in a consistant manner. First, the cathode bias was applied, establishing the drift field. Next, the EMCCD was set to record data. Once the EMCCD has began recording data, the THGEMs top electrode is biased to near their pre-established breakdown voltage, see Table~\ref{tab:maxBias}. This pre-established breakdown voltage was determined experimentally before the 24~hour discharge period, by slowing increasing the top electrode bias until THGEM discharges occured. 

Charging was investigated, through imaging alpha GAr interactions, over a 30~minute window. Care was taken to ensure that the position of the alpha source within the TPC was consistent between runs. By studying S2 light intensity variation over time, the dielectric charging behaviour of the THGEM can be inferred. After this 30~minute charging period, a THGEM bias scan characterisation study was undertaken.

The THGEM bias was reduced from near its predetermined breakdown voltage in 50~V intervals down to 1000~V (at this point the alpha tracks are barely discernible from the EMCCD sensor noise). For each THGEM bias, 2000 EMCCD images were recorded. Investigation of THGEM charging behaviour before the characterisation was a deliberate choice in order to maximally charge each THGEM before bias characterisation. The results of both experiments can be seen in Section~\ref{characterisation}.

\begin{table}[h!]
    \centering
    
    \begin{tabular}{|c|c|c|}
        \cline{1-3}
        \textbf{THGEM Type} & \textbf{Charging Study Bias (kV)} & \textbf{Breakdown Voltage (kV)}\\
    \hline
        FR4 circ & 1.60 & 1.80\\
        FS circ & 1.60 & 1.90 \\
        BF circ & 1.60 & 1.85\\
        BF hex & 1.70 & 1.80 \\
        \cline{1-3}
    \end{tabular}
    \caption{A Table of the THGEM bias during the charging test and the measured breakdown voltage, in 1100~mbar N6.0, flowing at a rate of 5L/min. The naming convention in the THGEM type is consistent with Table~\ref{tab:THGEMTypes}.}
    \label{tab:maxBias}
\end{table}

\subsection{GAr Purity Monitoring}
\label{Purity}
Purity was monitored using the well established $\tau_2$ lifetime method \cite{argonPurificationStudies}. Argon S1 VUV scintillation occurs through two decay paths: (singlet excimer) $\tau_1$ and (triplet excimer) $\tau_2$, the fast and slow component, respectively. As the $\tau_2$ decay time increases with argon purity, this value can be used as a relative purity monitor. It was important to ensure similar purity conditions for each G-THGEM/THGEM study as electronegative impurities, within GAr, could potentially vary the charge gain and S2 scintillation light production.

The highest values of $\tau_2$ lifetime reported for GAr is 3200$\pm$300~ns \cite{tau2Lifetime} (with variation depending on fitting models) and more recently 3480$\pm$65~ns \cite{Akashi-Ronquest2019_Tau2}. The PMT at the base of the TPC was used for all purity measurements. Purity measurements were taken after the initial fill of N6.0 GAr and subsequent measurements were taken while argon is continually flushed through the detector. Data collection began only once $\tau_2$ exceeded 2000~ns (which approximately corresponds to N6.0 argon gas \cite{argonPurificationStudies}).  A final $\tau_2$ lifetime measurement was taken at the end of each THGEM study, to validate that purity had been maintained throughout.

\section{THGEM Characteristics} \label{characterisation}
\begin{figure} [h!]
    \centering
    \includegraphics[scale=0.5]{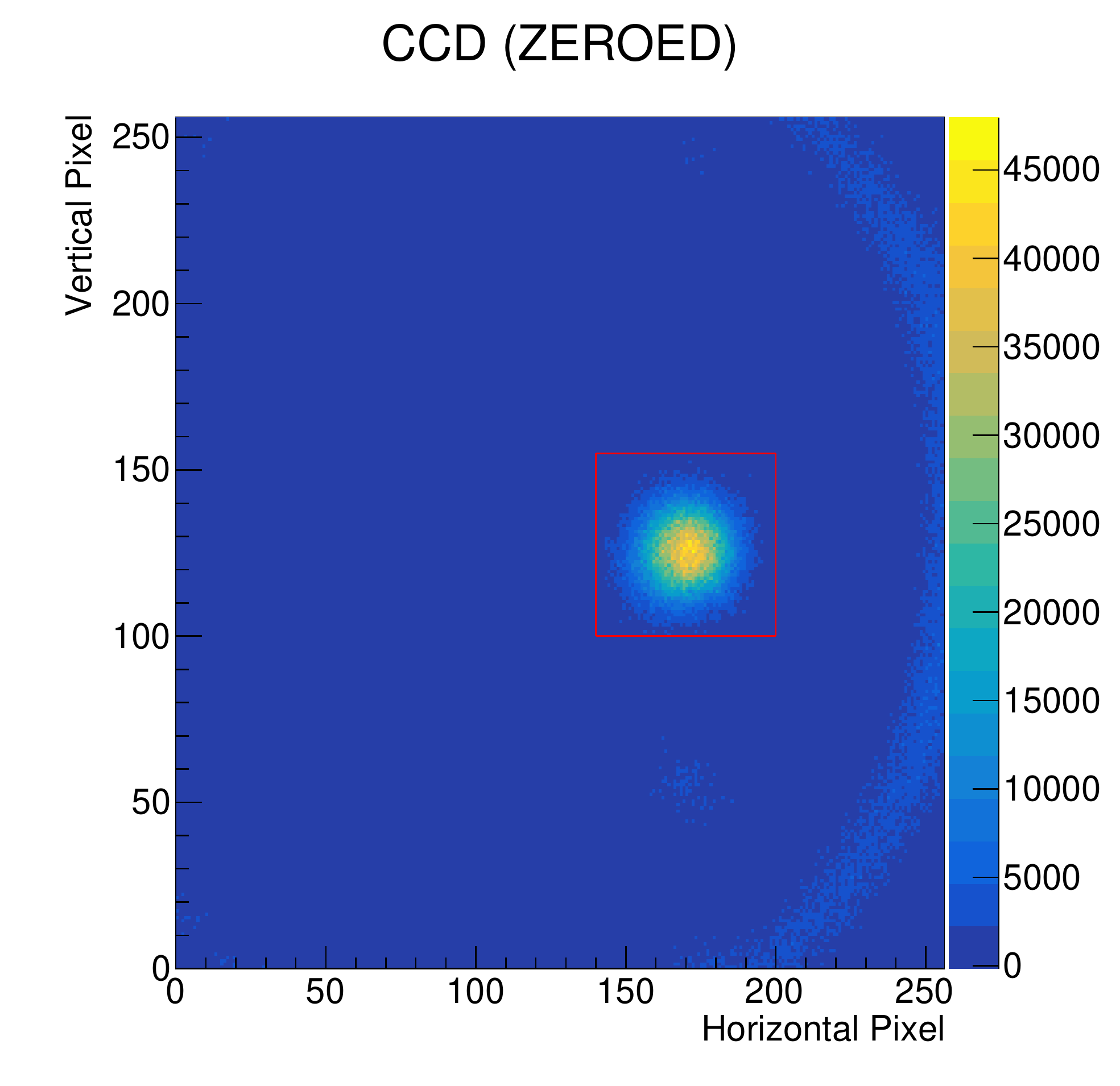}
    \caption{An EMCCD image of approximately 30 alpha particle GAr interactions, induced by a 30~kBq $^{241}$Am alpha source, collimated to 500~Bq. The EMCCD was configured in 4x4 binning, with a 50~ms exposure, and a gain of 1000. This was a direct image taken of the WLS S2 light produced within a Fused Silica substrate G-THGEM holes, at a field of 1600~V/mm. This image has been background subtracted. The red overlay is a ROI cut for analysis. }
    \label{fig:ROICut}
\end{figure}

Figure~\ref{fig:ROICut} shows an example of alpha particle interactions with GAr, as captured by the EMCCD. This image has been processed, with a background subtraction removing intrinsic sensor noise.  The background image was generated from data taken when the THGEM was fully discharged and unbiased. Subsequently, a Region Of Interest (ROI) cut was made which contains all of the alpha source S2 tracks - this ROI is highlighted in red on Figure~\ref{fig:ROICut}.

\subsection{Dielectric Charging Behaviour}

\begin{figure} [h!]
\centering
\begin{subfigure}{.48\textwidth}
  \centering
   \includegraphics[scale=0.4]{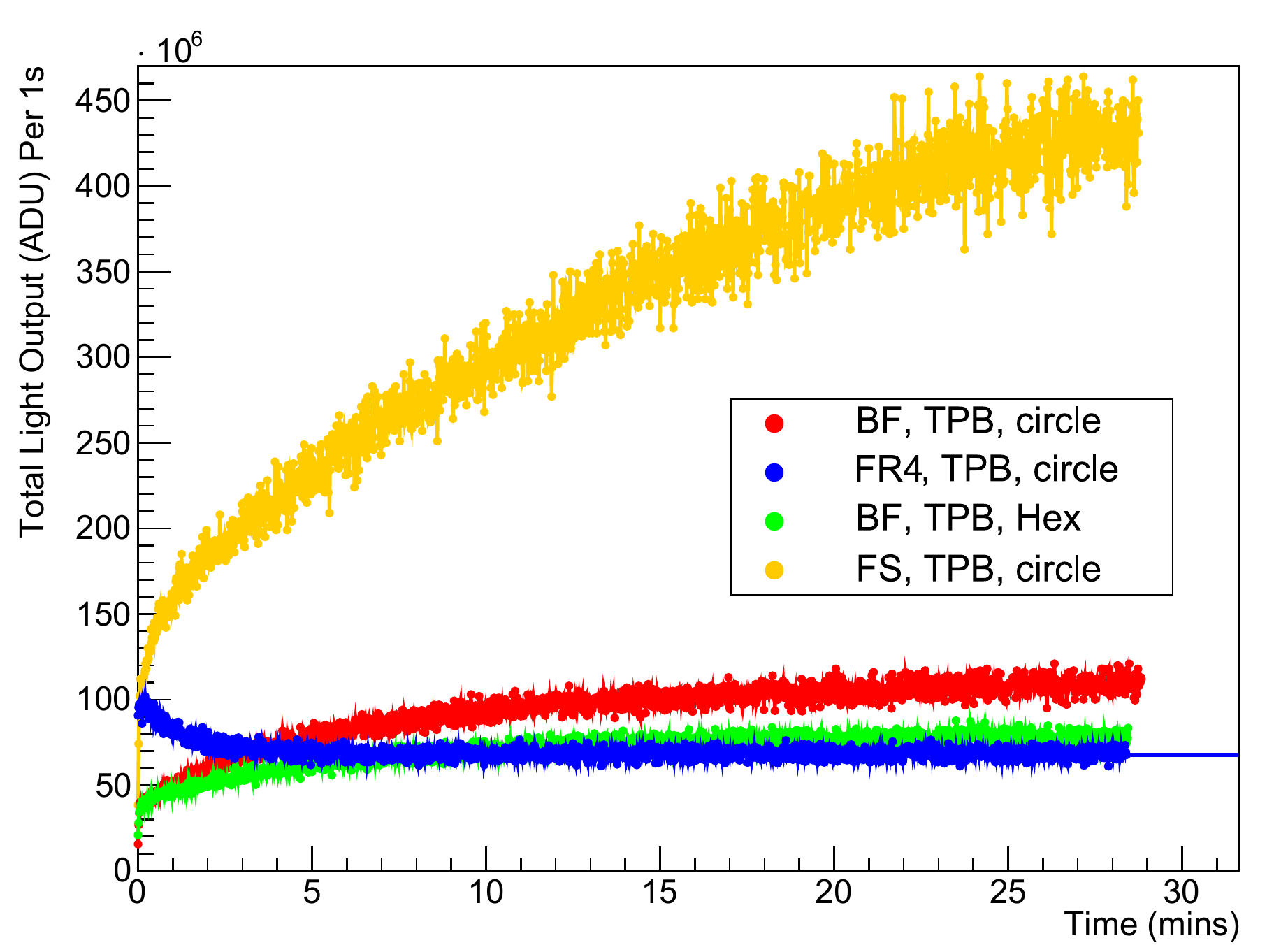}
  \caption{1s binning.}
  \label{fig:1sCharging}
\end{subfigure} %
\begin{subfigure}{.48\textwidth}
  \centering
   \includegraphics[scale=0.4]{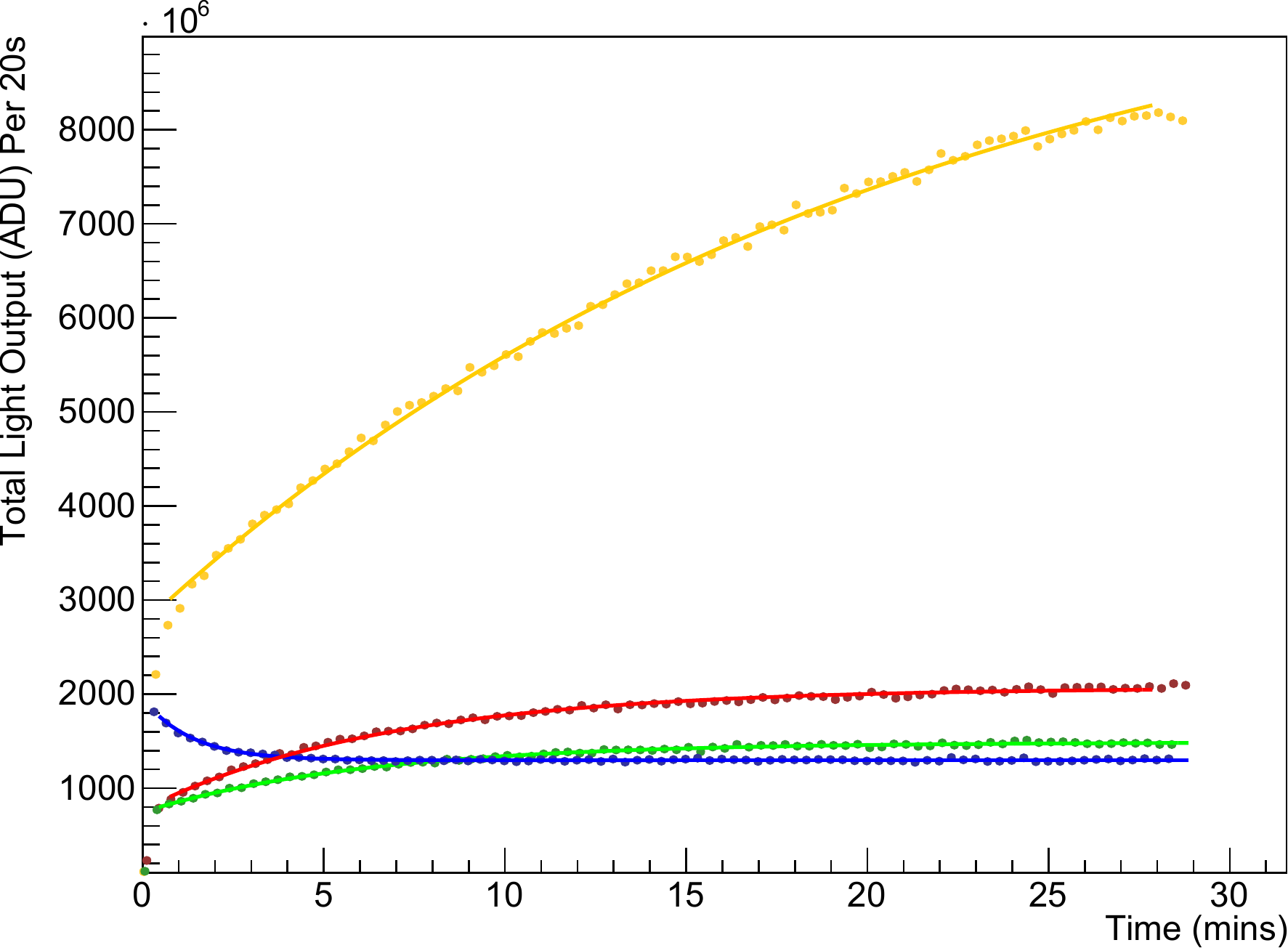}
  \caption{20s binning with fit.}
  \label{fig:20sCharging}
\end{subfigure}
\caption{The charging behaviour, manifesting as variation in S2 light intensity, of the various types of bi-conical hole G-THGEMs and cylindrical hole FR4 THGEM. Figures~\ref{fig:1sCharging} and \ref{fig:20sCharging} represent time binnings of 1~second and 20~seconds, respectively. All functions in Figure~\ref{fig:20sCharging} have been fitted using Equation~\ref{eq:Charging}, the parameters of which can be found in Table~\ref{tab:ChargingParameters}. The data colour is consistent with the legend on Figure~\ref{fig:charactersationPlots}.}
\label{fig:ChargingPlots}
\end{figure}

For each image, as detailed in Figure~\ref{fig:ROICut}, the pixel intensities inside the ROI were summed.  These summations were then binned into 1~second intervals shown in Figure~\ref{fig:1sCharging} and also 20~seconds  intervals shown in Figure~\ref{fig:20sCharging}. The shorter 1~second intervals better show gain fluctuations throughout the THGEM charging process whereas the longer 20~seconds intervals provide improved clarity for charging behaviour and shape.

The most apparent difference in charging behaviour between the G-THGEMs and the traditional FR4 THGEM is that the S2 light intensity of G-THGEMs increases over time, whereas the S2 light intensity of the conventional THGEM decreases over time. The distinct charging behaviours may be explained by differences in hole geometry. As a consequence of the abrasive machining process, G-THGEMs have bi-conical holes. The effect of bi-conical holes on dielectric charging behaviour is well understood for GEMs \cite{GEMChargingSim, GEMChargingCal}, and the theory can be extended to G-THGEMs. The insulating glass substrate provides a surface on which electrons may accumulate over time.  Consequently, this accumulated charge will distort the electric field within the G-THGEM hole. In the case of bi-conical holes, this accumulated charge causes an increase in electric field within the exit side of the hole (closer to the top electrode). This results in an increase in gain, and therefore increased light production. In contrast, THGEM holes with straight walls exhibit dielectric charging effects which result in the decrease of THGEM gain (and S2 light intensity) over time. This is primarily explained by a differential charge distribution along the walls of the holes \cite{THGEMChargingSim}.

All datasets shown on Figure~\ref{fig:20sCharging} are fitted with Equation~\ref{eq:Charging} \cite{THGEMChargingMeasurement, tripleGEMDetector}, where G is the total light intensity per 20~seconds. The parameters of Equation~\ref{eq:Charging} can be found in Table~\ref{tab:ChargingParameters}. The parameter $p_0$ is the initial gain, at time $T_0$. The direction and amplitude of charging is governed by $p_1$, where a negative (positive) $p_1$ value represents charging up (down), or more specifically, greater (less) S2 light production over time. The rate of charging is dictated by parameter $p_2$. 

\begin{equation}
   G = p_0+p_1e^{-t/p_2}
   \label{eq:Charging}
\end{equation}

\begin{table}[h!]
    \centering
    \small
        \setlength\tabcolsep{2pt}
    \begin{tabular}{|c|c|c|c|}
    \cline{1-4}
        \textbf{THGEM Type} & \textbf{$p_0$} & \textbf{$p_1$} & \textbf{$p_2$} \\
    \hline
        FR4 circ & $1.30\times 10^{9} \pm 1.26\times 10^{6}$ & $6.60\times 10^{8} \pm 1.42\times 10^{7}$ & 0.752 $\pm$ 0.0210 \\
        FS circ & $1.02\times 10^{10} \pm 2.36\times 10^{7}$ &  $-7.48\times 10^{9} \pm 4.12\times 10^{7}$ & 0.04808 $\pm$ 0.00560 \\
        BF circ & $2.07\times 10^{9} \pm 6.63\times 10^{6}$ & $-1.30\times 10^{9} \pm 1.23\times 10^{7}$ & 0.149 $\pm$ 0.00346\\
        BF hex & $1.49\times 10^{9} \pm 3.05\times 10^{6}$ & $-7.42\times 10^{8} \pm 6.81\times 10^{6}$ & 0.162 $\pm$ 0.00312 \\
        \cline{1-4}

    \end{tabular}
    \caption{A Table of the fit parameters for Equation~\ref{eq:Charging}, describing the fits in Figure~\ref{fig:20sCharging}.}
    \label{tab:ChargingParameters}
\end{table}

Despite similar bi-conical hole geometries, clear differences in dielectric charging behaviours are discernible between all G-THGEMs. The Borofloat 33 substrate G-THGEMs, with both circular and hexagonal hole geometries, had similar S2 light production at $T_0$. Additionally, both hole layouts resulted in maximal charging time, after around 10~mins. This is quantitatively reinforced by their $p_0$ and $p_2$ parameters of $2.07\times 10^{9}$ and 0.149 (circular) and $1.49\times 10^{9}$ and 0.162 (hexagonal), respectively. However, the larger hexagonal holes result in around half the light production in a maximally charged state than for the circular holes. The ratio of $p_1$ values dictates the relative total S2 light production differences at  maximal charging. Between circular and hexagonal hole layout, Borofloat 33 G-THGEMs, the $p_1$ ratio is 1.8. The reduction in S2 light intensity of the hexagonal hole G-THGEM can be attributed to a reduced field for larger holes at the same bias.

 Although both the Fused Silica and Borofloat 33 G-THGEMs have the same circular bi-conical hole geometries, a significant difference of the dielectric charging behaviour can be noticed in Figure~\ref{fig:ChargingPlots}. The ratio of the charging rate parameters, described by $p_2$, between Fused Silica and Borofloat 33 is 0.32, leading to the approximately observed charging rate of a third. The ratio of $p_1$ amplitude parameters is 5.8, describing the approximately six fold increase in amplitude after 30~mins of charging when comparing Fused silica to Borofloat 33. One possible explanation for these dielectric charging behaviour differences is that Borofloat 33 will typically contain many more impurities than Fused Silica. It is conceivable that these impurities may create a pathway by which electrons on the surface of the holes can move, thereby reducing the dielectric charging effect.

\subsection{S2 Light Production THGEM Bias Scan}

\begin{figure}  [h!]
    \centering
    \includegraphics[width=0.9\textwidth]{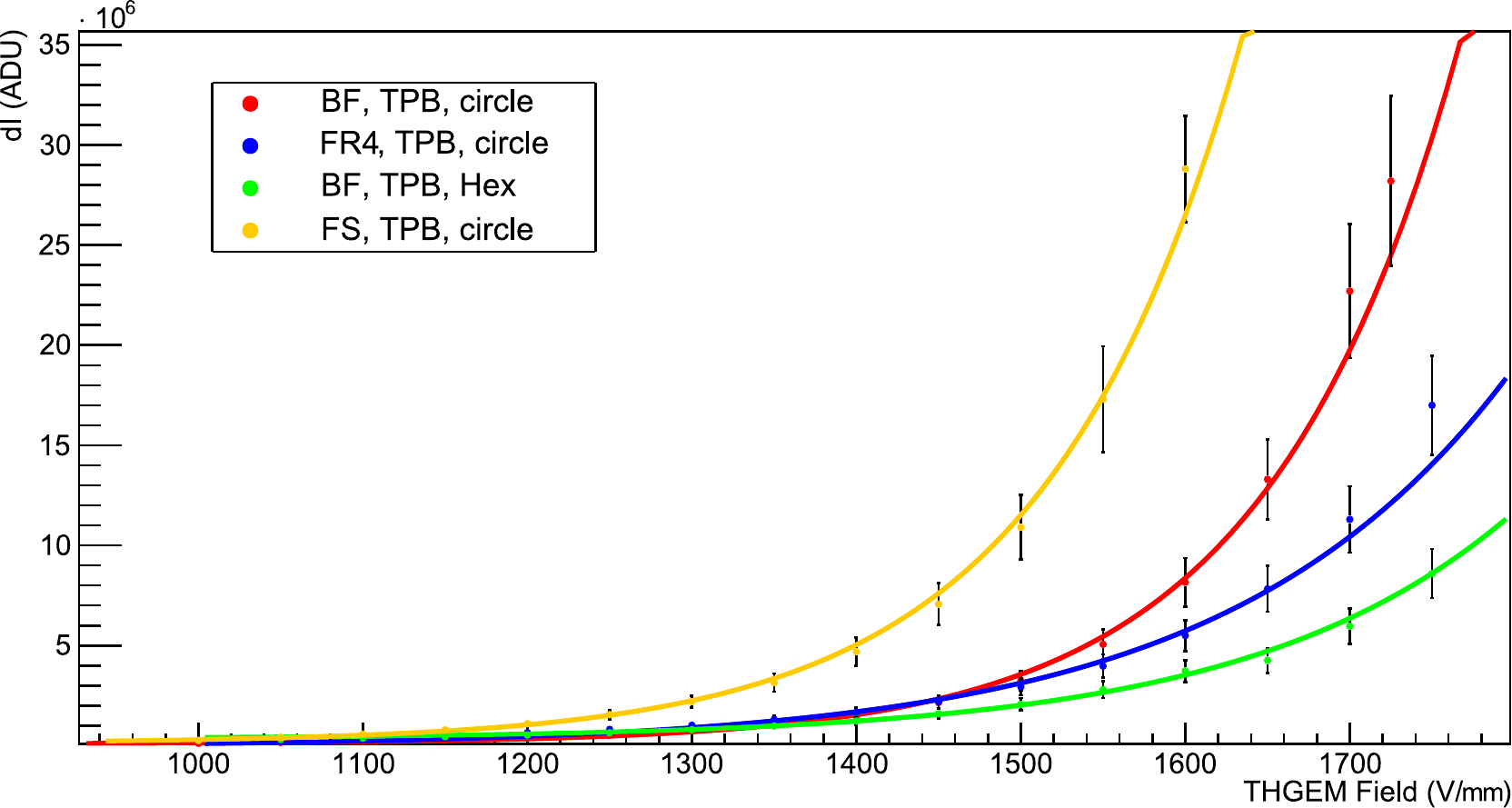}
    \caption{The FR4 THGEM and G-THGEM characterisation plots showing the total light intensity, at a variety of THGEM fields, for 2000 EMCCD images, each containing approximately 30 alpha tracks. All functions have been fitted with Equation~\ref{eq:Characterisation}. The parameters of the fits can be seen in Table~\ref{tab:CharacterisationParameters}.}
    \label{fig:charactersationPlots}
\end{figure}

A THGEM bias scan study was also performed, detailing the S2 light intensity as a function of THGEM field. Similarly to the method described for the dielectric charging behaviour, a ROI cut was taken around the alpha particles. This was done for all of the 2000 (background subtracted) EMCCD images collected for each THGEM type at a given bias. For each event, within this ROI, the pixel intensities were summed and a distribution produced of the total light intensity, in ADU, for the respective G-THGEMs/THGEM fields. The results of the study are shown in Figure~\ref{fig:charactersationPlots}. 

Nominal THGEM behaviour can be seen for both the G-THGEMs and the FR4 THGEM.  At low field (below approximately 1000V), the THGEMs are in the linear gain regime, the S2 light production is purely via electroluminescence.  As the field increases, the THGEM enters the electron multiplication regime, and the S2 light intensity is dominated by exponential effects.  The overall THGEM S2 light production behaviour can be described by Equation~\ref{eq:Characterisation} (a convolution of both linear and exponential regimes). The fit parameters for Equation~\ref{eq:Characterisation}, describing the data sets in Figure~\ref{fig:charactersationPlots}, can be seen in Table~\ref{tab:CharacterisationParameters}.

\begin{equation}
    I = p_0 x(1+p_1e^{p_2 x})+p_3
    \label{eq:Characterisation}
\end{equation}

\begin{table}[h!]
    \centering
    \small
        \setlength\tabcolsep{2pt}
    \hspace*{-2cm}\begin{tabular}{|c|c|c|c|c|}
    \cline{1-5}
        \textbf{THGEM Type} & \textbf{$p_0$} & \textbf{$p_1$} & \textbf{$p_2$} & \textbf{$p_3$} \\
    \hline
        FR4 circ & $29.7 \pm 8.0$ & $2.30\times 10^{-2} \pm 6.3\times 10^{-3}$ & $5.36\times 10^{-3} \pm 2.1\times 10^{-4}$ & $-7.68\times 10^{4} \pm 1.831\times 10^{5}$ \\
        FS circ & $117 \pm 57$ &  $5.50\times 10^{-4} \pm 2.47\times 10^{-4}$ & $7.78\times 10^{-3} \pm 3.5\times 10^{-4}$ & $2.97\times 10^{4} \pm 8.32\times 10^{4}$ \\
        BF circ & $7.15\times 10^{-3} \pm 3.59\times 10^{-3}$ & $1.85 \pm 3\times 10^{-2}$ & $8.05\times 10^{-3} \pm 3.0\times 10^{-4} $ & $9.30\times 10^{4} \pm 2.21\times 10^{4} $ \\
        BF hex & $8.37 \pm 5.97 $& $2.35\times 10^{-2} \pm 1.69\times 10^{-2}$ & $5.76 \times 10^{-3} \pm 4.0\times 10^{-4}$ & $3.40\times 10^{5} \pm 1.39\times 10^{5}$ \\
        \cline{1-5}

    \end{tabular}\hspace*{-2cm}
    \caption{A table of fit parameters for Equation~\ref{eq:Characterisation}, describing the fits in Figure~\ref{fig:charactersationPlots}.}
    \label{tab:CharacterisationParameters}
\end{table}

By far, the greatest light intensity comes from the Fused Silica G-THGEM. Both Borofloat 33 and Fused Silica G-THGEMs, which have hole sizes comparable to the FR4, produce increased S2 light intensity relative to the FR4 THGEM. This can be attributed to the distinct charging characteristics of the G-THGEMs. The Borofloat 33 G-THGEM with a hexagonal hole pattern produces the least S2 light intensity. This can be attributed to a larger hole, meaning that for the same bias the electric field within the hole is lower, therefore resulting in relatively less gain and thus reduced S2 light intensity. 

Both Fused Silica and Borofloat 33 G-THGEMs were found to be more resilient to damage than FR4, where irreparable carbonisation can occur after repeated discharges. As noted previously, carbonisation has the potential to severely degrade the performance of the THGEM. Nominally, the highly conducting copper electrode of classical THGEMs can allow a large current flow during discharges. The resistive nature of the ITO electrode coating of the G-THGEMs may possibly be quenching sparks by limiting current flow \cite{ResistiveElectrode}.

With regards to G-THGEM applications within TPCs, the benefits of increased S2 light intensity are obvious. Firstly, for the same applied bias, increased light intensity could possibly result in a lowered energy threshold of a TPC. Functioning at a lower bias, while giving comparable light intensity, reduces the possibility of discharges, both directly on the THGEM and on connectors, feedthroughs or cabling. Detector stability is an important consideration for long-term, large-scale experiments.

\section{Outlook and Future Developments}
\label{conclusion}
The overarching aim of this investigation was the assessment of a new manufacturing process for G-THGEMs, in the context for applications within large-scale LArTPCs, and potentially beyond into dark matter and medical research. Within this paper, three G-THGEMs were produced using a novel, patent pending, masked abrasive machining technique. This manufacturing method gives unprecedented versatility to the substrates and electrodes materials, together with hole shape and pattern. For first characterisation (and comparison with an FR4 THGEM), a selection of G-THGEMs were produced with unique behaviours. Variations in glass substrate (either Borofloat 33 or Fused Silica) were studied, as well as two distinct hole patterns (circular or hexagonal). A transparent ITO electrode was used for all G-THGEMs. 

General reported issues with current THGEM technologies in LArTPC applications, can be summarised as the following: material bowing, limited stiffness and substrate thickness variations all resulting in field non-uniformities, contamination due to the porous nature of FR4, and discharges affecting long term stability. In addition, secondary problems arise when considering quality assurance for the scale of production required for a kiloton scale experiment, such as DUNE - in particular, selection of FR4 substrates with suitable thickness variations and consistent hole formation from mechanical drilling. Furthermore, current FR4 THGEMs are unsuitable for use within the dark matter community due to inadequate substrate radiopurity. G-THGEMs have the potential to alleviate many of these concerns, while maintaining quality control and quality assurance  manufacturing required for large-scale experiments.

Characterisation of the G-THGEMs (compared to a conventionally manufactured FR4 THGEM) determined that for the same hole configuration, at the same bias, G-THGEMs have increased S2 light production. Furthermore, the bi-conical shape of the G-THGEM holes produced by the abrasive machining process leads to an increasing S2 light intensity over time as the G-THGEMs charge up. Cylindrical holes, typical of conventional THGEMs, result in the decrease of light intensity over time.

The masked abrasive machining process allows for unprecedented customisation of G-THGEMs. The substrate, electrode, hole shape, size and pattern can all be varied, selecting for certain desirable mechanical and behavioural properties. For example, within this experiment, Borofloat 33 and Fused Silica glass were both investigated. Borofloat 33 is relatively low-cost, whereas Fused Silica is radiopure and may have the potential for use within dark matter TPCs. Both of these glasses boast flexural moduli of between 64 and 72~GPa, making them almost 3 times stiffer than FR4 (23~GPa). In addition, glass substrates are available with thickness variations better than $<1.5\%$, a marked improvement on the typically achieved $5\%$ for FR4 substrate THGEMs. Finally, the observed discharge behaviour of G-THGEMs was different to typical THGEMs. As ITO is resistive, discharges are localised to the region and the amount of current flow is limited. This has an effect of quenching sparks, potentially reducing damage to the device. Additionally, carbonisation, a known problem with FR4, does not occur with glass substrates. 

The manufacturing technique and results discussed within this paper represents an encouraging first demonstration of the feasibility of masked abrasively machined G-THGEMs. While still in their infancy G-THGEMs, are a promising avenue for exploration. The abrasive machining technique allows for unprecedented versatility in THGEM design, allowing configurations suitable for a range of applications. Further development is ongoing, including the production of a larger G-THGEMs for incorporation into dual-phase LArTPCs. Specifically, production will include larger $54 \times 54$~cm$^2$ G-THGEMs for use in the ARIADNE detector, as well as sixteen $50 \times 50$~cm$^2$ G-THGEMs for large scale demonstrations at the CERN Neutrino Platform \cite{Amedo2020_LOI}. 

\section{Patents}

K. Mavrokoridis, A. Roberts and The University of Liverpool, 2020, Gas Electron Multiplier, Patent Pending, GB2019563.2

\vspace{6pt} 



\authorcontributions{Conceptualization, K.Mv., B.P., A.R. and C.T.; Funding acquisition, K.Mv. and C.T.; Investigation, A.L., K.Mj., K.Mv., B.P. and A.R.; Writing, A.L., K.Mj., K.Mv., B.P., A.R. and C.T. All authors have read and agreed to the published version of the manuscript.}

\funding{The ARIADNE program is funded by the European Research Council Grant No. 677927.}

\acknowledgments{The authors would like to thank the members of the Mechanical Workshop of the University of Liverpool's Physics Department, for their contributions and invaluable expertise.}

\conflictsofinterest{The authors declare no conflict of interest. The funders had no role in the design of the study; in the collection, analyses, or interpretation of data; in the writing of the manuscript, or in the decision to publish the results.}

\reftitle{References}



\begin{thebibliography}{999}


\bibitem{GEM}
Sauli F.,
GEM: A new concept for electron amplification in gas detectors,
Nuclear Instruments and Methods in Physics Research Section A: Accelerators, Spectrometers, Detectors and Associated Equipment,
Volume 386, Issues 2–3,
1997,
Pages 531-534.

\bibitem{THGEM}
Chechik, R.; Breskin, A.; Shalem, C.; Mormann, D.,
Thick GEM-like hole multipliers: properties and possible applications,
Nucl. Instrum. Meth. A,
Volume 535, Issue 1-2 (2004), Pages 303-308

\bibitem{THGEMDetectors}
Breskin, A.; Alon, R.; Cortesi, M.; Chechik, R.; Miyamoto, J.; Dangendorf, V.; Maia, J.; Dos Santos, J. M. F., 
A concise review on THGEM detectors, 
Nucl. Instrum. Meth. A,
Volume 598, (2009), Pages 107-111

\bibitem{Buzulutskov2020_ReviewChargeAmplification}
Buzulutskov, A. Electroluminescence and Electron Avalanching in Two-Phase Detectors. Instruments 2020, 4, 16. 
\url{https://doi.org/10.3390/instruments4020016}

\bibitem{DuneCDRVol1}
DUNE Collaboration, Long-Baseline Neutrino Facility (LBNF) and Deep Underground Neutrino Experiment (DUNE) Conceptual Design Report Volume 1: The LBNF and DUNE Projects [arXiv:1601.05471]

\bibitem{DuneCDRVol2}
DUNE Collaboration, Long-Baseline Neutrino Facility (LBNF) and Deep Underground Neutrino Experiment (DUNE) Conceptual Design Report Volume 2: The Physics Program for DUNE at LBNF [arXiv:1512.06148]

\bibitem{DuneCDRVol3}
Strait, J.; McCluskey, E.; Lundin, T.; Willhite, J.; Hamernik, T.; Papadimitriou, V.; Marchionni, A. ; Kim, M. J.; Nessi, M.; Montanari, D.; Heavey, A., 
Long-Baseline Neutrino Facility (LBNF) and Deep Underground Neutrino Experiment (DUNE) Conceptual Design Report Volume 3: Long-Baseline Neutrino Facility for DUNE June 24, 2015 [arXiv:1601.05823]

\bibitem{DuneCDRVol4}
DUNE Collaboration, Long-Baseline Neutrino Facility (LBNF) and Deep Underground Neutrino Experiment (DUNE) Conceptual Design Report, Volume 4 The DUNE Detectors at LBNF [arXiv:1601.02984]

\bibitem{DuneIDRVol1}
DUNE Collaboration, The DUNE Far Detector Interim Design Report Volume 1: Physics, Technology and Strategies, [arXiv:1807.10334]

\bibitem{DuneIDRVol2}
DUNE Collaboration, The DUNE Far Detector Interim Design Report, Volume 2: Single-Phase Module, [arXiv:1807.10327]

\bibitem{DuneIDRVol3}
DUNE Collaboration, The DUNE Far Detector Interim Design Report, Volume 3: Dual-Phase Module, [arXiv:1807.10340]

\bibitem{CYGNO}
Baracchini, E.; Benussi, L.; Bianco, S.; Capoccia, C.; Caponero, M.; Cavoto, G.; Cortez, A.; Costa, I. A.; Di Marco, E.; D'Imperio, G.; et al.,
CYGNO: a gaseous TPC with optical readout for dark matter directional search,
Journal of Instrumentation,
Volume 15, Issue 7 (2020), Pages C07036

\bibitem{Darkside}
Aalseth, C. E.; Acerbi, F.; Agnes, P.; Albuquerque, I. F. M.; Alexander, T.; Alici, A.; Alton, A. K.; Antonioli, P.; Arcelli, S.; Ardito, R.; et al.,
DarkSide-20k: A 20 Tonne Two-Phase LAr TPC for Direct Dark Matter Detection at LNGS,
Eur. Phys. J. Plus,
Volume 133, (2018), Pages 131

\bibitem{ArDM}
Marchionni, A.; Amsler, C.; Badertscher, A.; Boccone, V.; Bueno, A.; Carmona-Benitez, M. C.; Coleman, J.; Creus, W.; Curioni, A.; Daniel, M.; et al.,
ArDM: a ton-scale LAr detector for direct Dark Matter searches,
2011 J. Phys.: Conf. Ser. 308 012006

\bibitem{LZ}
Mount, B. J.; Hans, S.; Rosero, R.; Yeh, M.; Chan, C.; Gaitskell, R. J.; Huang, D. Q.; Makkinje, J.; Malling, D. C.; Pangilinan, M.; et al., 
LUX-ZEPLIN (LZ) Technical Design Report, 2017
[arXiv:1703.09144]

\bibitem{Mavrokoridis2020_G-THGEM}
Mavrokoridis, K., Roberts, A., and The University of Liverpool, 2020, Gas Electron Multiplier, Patent Pending, GB2019563.2. 

\bibitem{ARIADNE_TDR}
Hollywood, D.; Majumdar, K.; Mavrokoridis, K.; McCormick, K. J.; Philippou, B.; Powell, S.; Roberts, A.; Smith, N. A.; Stavrakis, G.; Touramanis, C.; Vann, J.,
ARIADNE - A Novel Optical LArTPC: Technical Design Report and Initial Characterisation using a Secondary Beam from the CERN PS and Cosmic Muons,
Journal of Instrumentation,
Volume 15, Issue 3 (2020), Page P03003

\bibitem{Takahashi_2013}
Takahashi, H.; Mitsuya, Y.; Fujiwara, T.; Fushie, T.,
Development of a glass GEM,
Nucl. Instrum. Meth. A,
Volume 724, Issue 1 (2013), Pages 1-4

\bibitem{Fujiwara_2019}
Fujiwara, T.; Mitsuya, Y.;  Fushie, T.; Aoki, T.,
Demonstration of soft X-ray 3D scanning and modeling with a glass gas electron multiplier,
Journal of Instrumentation,
Volume 14, (2019), Page P11022

\bibitem{Gai_2007_DarkMatterTHGEMs}
Gai, M.; Alon, R.; Breskin, A.; Cortesi, M.; McKinsey, D. N.; Miyamoto, J.; Ni, K.; Rubin, D. A. R.; Wongjirad, T.,
Toward Application of a Thick Gas Electron Multiplier (THGEM) Readout for a Dark Matter Detector
[arXiv:0706.1106]

\bibitem{Tsyganov_2008_Medical}
Tsyganov, E.; Antich, P.; Parkey. R.; Seliounine, S.; Golovatyuk, V.; Lobastov, S.; Zhezher, V.; Buzulutskov, A.; 
Gas Electron Multiplying Detectors for medical applications
Nucl. Instrum. Meth. A,
Volume 597, Issue 2-3, (2008), Page 257-265



\bibitem{Abgrall2016}
Abgrall, N.; Arnquist, I. J.; Avignone, F. T.; Back, H. O.; Barabash, A. S.; Bertrand, F. E.; Boswell, M.; Bradley, A. W.; Brudanin, V.; Busch, M.; et al.,
The Majorana Demonstrator radioassay program,
Nuclear Instruments and Methods in Physics Research Section A: Accelerators, Spectrometers, Detectors and Associated Equipment,
Volume 828,
2016,
Pages 22-36

\bibitem{LaserEtching}
Tanaka, R.; Takaoka, T.; Mizukami, H.; Arai, T.; Iwai, Y., Laser etching of indium tin oxide thin films by ultra-short pulsed laser, Proc. SPIE 5063, Fourth International Symposium on Laser Precision Microfabrication, (18 November 2003); \url{https://doi.org/10.1117/12.540533}


\bibitem{SCHOTT_BF}
SCHOTT BOROFLOAT 33 – Mechanical Properties, 
\url{https://www.schott.com/d/borofloat/723d30c8-cca0-4159-ad40-31e658dbf588/1.8/borofloat33_mech_eng_web_09_2020.pdf}
(Accessed: 2021/02/21)

\bibitem{borofloatData}
Borofloat 33 Data Sheet,  \url{https://abrisatechnologies.com/products-services/glass-products/borosilicate/schott-borofloat-33/}
(Accesssed: 2020/09/11)

\bibitem{uqgoptics}
UQG Optics SCHOTT Borofloat 33, 
\url{https://www.uqgoptics.com/wp-content/uploads/2019/07/Schott-Borofloat-33.pdf}
(Accessed: 2021/07/21)

\bibitem{SCHOTT_TechnicalGlasses}
SCHOTT Technical Glasses, 
\url{https://www.schott.com/d/uk/76a1e227-63ac-4f88-b111-efd65c41c95e/1.0/schott_techn_glaeser_e_2007.pdf}
(Accessed: 2021/02/21)

\bibitem{SCHOTT_FS}
SCHOTT Synthetic Fused Silica, 
\url{http://www.jmcglass.com/down/fused_silica_us.pdf}
(Accessed: 2021/02/21)

\bibitem{fusedSilicaData}
Fused Silica Data Sheet, 
\url{https://abrisatechnologies.com/products-services/glass-products/quartz-fused-silica/corning-7980/}
(Accesssed: 2020/09/11)

\bibitem{SCHOTT_GlassWafers}
SCHOTT Glass Wafers Specifications 
\url{https://www.schott.com/d/advanced_optics/7af2454e-555a-4071-9590-bb7b32e75fec/schott-glass-wafer-specification-english-26062018.pdf}
(Accessed: 2021/09/03)

\bibitem{Eltos_Materials}
Eltos Material Portfolio, 
\url{http://www.eltos.com/en/documents/}
(Accessed: 2021/07/21)

\bibitem{Isola_DE104}
Isola Standard DE104 FR4, 
\url{https://www.isola-group.com/pcb-laminates-prepreg/de104/}
(Accessed: 2021/05/05)

\bibitem{ESA}
European Space Agency ESMAT, 
\url{http://esmat.esa.int/index.html}
(Accessed:2021/07/28)

\bibitem{ResistiveElectrode}
Song, G.; Zhou, Y.; Shao, M.; Shang, L.; Lv, Y.; Wang, X.; Liu, J.; Zhang, Z.,
Development of THGEM-like detectors with diamond-like carbon resistive electrodes,
Journal of Instrumentation, 
Volume 15, November 2020, P11013

\bibitem{ARIADNE_Prototype}
Mavrokoridis, K.; Carroll, J.; McCormick, K. J.; Paudyal, P.; Roberts, A.; Smith, N. A.; Touramanis, C.,
First Demonstration of Imaging Cosmic Muons in a Two-Phase Liquid Argon TPC using an EMCCD Camera and a THGEM,
Journal of Instrumentation, 
Volume 10, 2015, P10004

\bibitem{PMTDataSheet}
Hamamatsu R5912-20 PhotoMultipler Tube Data Sheet. \url{https://www.hamamatsu.com/resources/pdf/etd/LARGE_AREA_PMT_TPMH1376E.pdf}
(Accesssed: 2020/01/04)

\bibitem{ArgonScintillation}
Heindl, T.; Dandl, T.; Hofmann, M.; Krucken, R.; Oberauer, L.; Potzel, W.; Wieser, J.; Ulrich, A., 
The scintillation of liquid argon, 
Europhysics Letters,
Volume 91, Issue 6, 2010, Pages 62002

\bibitem{TPBSpectrum}
Gehman, V. M.; Seibert, S. R.; Rielage, K.; Hime, A.; Sun, Y.; Mei, D. M.; Maassen, J.; Moore, D.,
Fluorescence efficiency and visible re-emission spectrum of tetraphenyl butadiene films at extreme ultraviolet wavelengths,
Nucl. Instrum. Meth. A,
Volume 654, Issue 1 (2011), Pages 116-121

\bibitem{Benson2017_TPBSpectrum}
Benson, C. P.; Orebi Gann, G. D.; Gehman, V. M., Measurements of the intrinsic quantum efficiency and absorption length of tetraphenyl butadiene thin films in the vacuum ultraviolet regime,
Eur.Phys.J.C 78 (2018) 4, 329

\bibitem{AndorDataSheet}
Andor - iXon Ultra 888 Specifications,
\url{http://www.andor.com/pdfs/specifications/iXon_Ultra_888_EMCCD_Specifications.pdf} (Accessed: 2020/01/04)

\bibitem{ViewportDataSheet}
Standard 7056 Kodial (Borosilicate) Glass Viewports
\url{https://www.lewvac.co.uk/product/standard-7056-kodial-borosilicate-glass/}
(Accessed: 2021/05/10)

\bibitem{THGEMReadout}
Lightfoot, P. K.; Barker, G. J.; Mavrokoridis, K.; Ramachers, Y. A.; Spooner, N. J. C.,
Optical readout tracking detector concept using secondary scintillation from liquid argon generated by a thick gas electron multiplier,
Journal of Instrumentation,
Volume 4, Issue 4 (2009), Page P04002

\bibitem{THGEMLightYield}
Monteiro, C. M. B.; Fernandes, L. M. P.; Veloso, J. F. C. A.; Oliveira, C. A. B.; dos Santos, J. M. F.,
Secondary scintillation yield from GEM and THGEM gaseous electron multipliers for direct dark matter search,
Physics Letters B,
Volume 714, Issue 1 (2012), Pages 18-23

\bibitem{THGEMReview_Buzulu}
Buzulutskov, A.,
Electroluminescence and Electron Avalanching in Two-Phase Detectors,
MDPI Instruments,
Volume 4, Issue 2 (2020), Article Number 16


\bibitem{argonPurificationStudies}
Mavrokoridis, K.; Calland, R. G.; Coleman, J.; Lightfoot, P. K.; McCauley, N.; McCormick, K. J.; Touramanis, C., 
Argon Purification Studies and a Novel Liquid Argon Re-circulation System,
Journal of Instrumentation,
Volume 6, 2011, P08003

\bibitem{tau2Lifetime}
Keto, J. W.; Gleason, R. E.; Walters, G. K., 
Production Mechanisms and Radiative Lifetimes of Argon and Xenon Molecules Emitting in the Ultraviolet, 
Phy. Rev. Lett. 33
(1974) 1365

\bibitem{Akashi-Ronquest2019_Tau2}
Akashi-Ronquest, M.; Bacon, A.; Benson, C.; Bhattacharya, K.; Caldwell, T.; Formaggio, J. A. ; Gastler, D.; Grado-White, B.; Griego, J.; Gold, M.; et al., Triplet lifetime in gaseous argon. Eur. Phys. J. A 55, 176 (2019). \url{https://doi.org/10.1140/epja/i2019-12867-2}


\bibitem{THGEMChargingMeasurement}
Pitt, M.; Correia, P. M. M.; Bressler, S.; Coimbra, A. E. C.; Renous, D. S.; Azevedo, C. D. R.; Veloso, J. F. C. A.; Breskin, A., 
Measurements of charging-up processes in THGEM-based particle detectors,
Journal of Instrumentation,
Volume 13, 2018, P03009

\bibitem{tripleGEMDetector}
Chatterjee, S.; Sen, A.; Roy, S.; G, K. Nivedita; Paul, A.; Das, S.; Biswas, S.,
Study of charging up effect in a triple GEM detector,
Journal of Instrumentation,
Volume 15, 2020, Pages T09011

\bibitem{GEMChargingSim}
Alfonsi, M.; Croci, G.; Duarte Pinto, S.; Rocco, E.; Ropelewski, L.; Sauli, F.; Veenhof, R.; Villa, M.,
Simulation of the dielectric charging-up effect in a GEM detector,
Nucl. Instrum. Meth. A,
Volume 671, 11 April 2012, Pages 6-9.

\bibitem{GEMChargingCal}
Correia, P. M. M.; Oliveira, C. A. B.; Azevedo, C. D. R.; Silva, A. L. M.; Veenhof, R.; Varun Nemallapudi, M.; Veloso, J. F. C. A., 
A dynamic method for charging-up calculations: the case of GEM,
Journal of Instrumentation,
Volume 9, 2014, P07025

\bibitem{THGEMChargingSim}
Correia, P. M. M.; Pitt, M.; Azevedo, C. D. R.; Breskin, A.; Bressler, S.; Oliveira, C. A. B.; Silva, A. L. M.; Veenhof, R.; Veloso, J. F. C. A., 
Simulation of gain stability of THGEM gas-avalanche particle detectors,
Journal of Instrumentation,
Volume 13, 2018, P01015

\bibitem{Amedo2020_LOI}
 Amedo, P.; Gonzalez-Dıaz, D.; Lowe, A.; Majumdar, K.; Mavrokoridis, K.; Nessi, M.; Philippou, B.; Pietropaolo, F.; Resnati, F.; Roberts, A.; Saa, A.; Touramanis, C.; Vann, J.,
 Letter  of  Intent:   Large-scale demonstration  of  the  ARIADNE  LArTPC  optical  readout  system  at the CERN Neutrino Platform,
 CERN, Geneva, Tech. Rep., Oct 2020.[Online], CERN-SPSC-2020-026, SPSC-I-255. 
 Available:  \url{http://cds.cern.ch/record/2739360}

\end{thebibliography}
\end{document}